\documentclass[12pt]{article}

\usepackage{newtxtext,newtxmath}

\usepackage{graphicx}

\usepackage[letterpaper,margin=1in]{geometry}

\renewenvironment{abstract}
	{\quotation}
	{\endquotation}

\date{}

\makeatletter
\renewcommand{\fnum@figure}{\textbf{Figure \thefigure}}
\renewcommand{\fnum@table}{\textbf{Table \thetable}}
\makeatother

\usepackage{scicite}

\usepackage{url}

\newcommand{\araa}{Annu. Rev. Astron. Astrophys.}   
\newcommand{\aj}{Astron. J.}   
\newcommand{\apj}{Astrophys. J.}   
\newcommand{\apjl}{Astrophys. J. Lett.}   
\newcommand{\apjs}{Astrophys. J. Suppl. Ser.}   
\newcommand{\aap}{Astron. Astrophys.}   
\newcommand{\aapr}{Astron. Astrophys. Rev.}   
\newcommand{\mnras}{Mon. Not. R. Astron. Soc.}   
\newcommand{\nat}{Nature} 
\newcommand{\pasj}{Publ. Astron. Soc. Jpn}   
\newcommand{\pasp}{Publ. Astron. Soc. Pac.}   

\def\scititle{
Stellar feedback drives the baryon deficiency in low-mass galaxies 
}
\title{\bfseries \boldmath \scititle}

\author{
	Haoran~Yu$^{1,2}$,
	Enci~Wang$^{1,2\ast}$,
    Zeyu~Chen$^{1,2}$,
    C\'eline~P\'eroux$^{3,4}$,\and
    Hu~Zou$^{5}$,
    Zhicheng~He$^{1,2}$,
    Huiyuan~Wang$^{1,2}$,
    Cheqiu~Lyu$^{1,2}$,\and
    Cheng~Jia$^{1,2}$,
    Chengyu~Ma$^{1,2}$,
	Xu~Kong$^{1,2}$\and
    \small$^{1}$Department of Astronomy, University of Science and Technology of China,\\ \small Hefei 230026, People's Republic of China\and
    \small$^{2}$School of Astronomy and Space Science, University of Science and Technology of China,\\ \small Hefei 230026, People's Republic of China\and
    \small$^{3}$European Southern Observatory, Karl-Schwarzschild-Str. 2, \\ \small 85748 Garching-bei-M\"unchen, Germany\and
    \small$^{4}$Aix Marseille Universit\'e, CNRS, LAM (Laboratoire d'Astrophysique de Marseille) \\ \small UMR 7326, 13388, Marseille, France\and
    \small$^{5}$National Astronomical Observatories, Chinese Academy of Sciences,\\ \small Beijing 100012, People's Republic of China\and
	\small$^\ast$Corresponding author. Email: ecwang16@ustc.edu.cn
}

\begin{document} 

\maketitle

\noindent
{\bf Short title:} Feedback drives baryon out of low-mass galaxies\\
{\bf Teaser:} Intense star formation drives powerful outflows, ejecting vast masses of baryonic material from low-mass dark matter halos

\subsection*{Abstract}
\begin{abstract} \bfseries \boldmath
Stellar feedback, as a key process regulating the baryon cycle, is thought to greatly redistribute baryonic material inside and outside the dark matter halos (DMHs), however the observational evidences are lacking.
Through stacking analyses of $\sim400,000$ galaxy spectra from Dark Energy Spectroscopic Instrument (DESI), we find star formation driven cool outflows in Mg\,\textsc{ii} absorption line.
Assuming only gravity acts on the launched gas, our calculations reveal that outflows from low mass galaxies ($M_*<10^{10}\,\rm M_\odot$) are capable of escaping beyond the DMHs, which aligns well with our finding in the circumgalactic medium (CGM) absorption along the minor-axes of galaxies using background quasars.
This research offers indirect evidence that stellar feedback drives the low baryon retention rate in low-mass halos, implicating that baryonic processes within galaxies are connected with the diffuse matter beyond the DMHs.
\end{abstract}

\noindent

\subsection*{Introduction}

Star-forming galaxies undergo intense baryonic processes within their interiors, interacting with their surrounding environments during their lifetimes.
These processes include gas accretion onto DMHs and galaxies~\cite{vandeVoort11}, gas cooling and collapse due to gravitational instability~\cite{Thompson16}, gas consumption during star formation, and stellar feedback in the form of stellar winds and supernova explosions, driving outflows in galactic scale~\cite{Veilleux05, Veilleux20, Thompson24}.
In numerical simulations of galaxy formation and evolution, gas ejection launched by stellar feedback is introduced to explain the low baryon retention fraction in low-mass galaxies~\cite{Behroozi10, Moster13, Puchwein13, Vogelsberger20}.
However, hindered by the limited information of the multi-phase baryonic matter from observations, no clear evidence has justified the connection between the baryon deficiency in galaxies and the feedback processes.
In the meantime, the physical mechanisms of feedback processes are interpreted quite differently in various suits of simulations~\cite{Hopkins14,Hopkins18,Schaye15,Vogelsberger14,Nelson15}.
The critical observational evidence is of urgent need, potentially challenging the theory of galaxy evolution.

In this work, we study the cool outflows in star-forming galaxies through utilizing tremendous amount of galaxy spectra from the early data release of Dark Energy Spectroscopic Instrument~\cite{DESIedr} (DESI EDR).
In the so-called ``down-the-barrel'' methodology, the foreground medium imprints absorption features on the galaxy spectrum, enabling its kinematics to be investigated via the Doppler effect.
Through spectral stacking analysis, we derive statistically robust composite spectra for galaxies of different stellar masses ($M_*$) and star-formation rates ($\rm SFR$).
We find cool outflows prevalent in star-forming galaxies in the form of blueshifted absorption line of singly ionized magnesium (Mg\,\textsc{ii}).
Through combined analysis of Mg\,\textsc{ii} absorption in the CGM surrounding those galaxies using background quasars (QSOs), we provide robust observational evidence, based on an unprecedented dataset, that stellar feedback drives sufficiently strong outflows to account for the baryon deficiency in low-mass galaxies.

\subsection*{Results}

Analogous to the main sequence in the Hertzsprung–Russell diagram, which is essential in the study of stellar evolution, the Star-Forming Main Sequence (SFMS)~\cite{Brinchmann04, Noeske07, Speagle14} --- a fundamental correlation between $M_*$ and $\rm SFR$ --- serves as a critical framework for investigating galaxy evolution.
We stack galaxy spectra based on their positions within the SFMS, as shown in the left panel of Figure~\ref{fig:map}.
The samples are binned into hexagonal regions with a width length of 0.05 dex, while the peripheral ones enclosing fewer than 500 galaxies are excluded.
The color of each hexagon corresponds to the amount of spectra within that bin and its adjacent neighbors, which are combined to obtain the stacked spectra.
With most bins containing more than 10,000 spectra, the resulting stacked spectra achieve exceptionally high signal-to-noise ratio ($\rm S/N\gtrsim100$), as demonstrated in the inset. 
The stacked profile of Mg\,\textsc{ii} doublet exhibits a near-unity ratio, in contrast to the 2:1 ratio expected in the optically thin limit~\cite{Morton03}. 
This behavior, consistent with previous individual and stacked down-the-barrel studies~\cite{Erb12, Rubin14, Perrota23, Pessa24, Valentino25}, implies optically thick, saturated absorption.

Using the {\sc Boxcar} method~\cite{Rubin10, Bordoloi14}, we extract the outflow equivalent width ($\rm EW_{out}$) and mean outflow velocity ($v_{\rm out}$) of the Mg\,\textsc{ii} absorption features from the composite spectra, as shown in the top-right panels of Figure~\ref{fig:map}.
Derived directly from the stacked spectra, $\mathrm{EW}_{\rm out}$ and $v_{\rm out}$ characterize the Mg~\textsc{ii} absorption and serve as the primary results of this study, with minimal dependence on model assumptions.
Assuming the geometry of the outflow to be an expanding spherical thin shell launching at $5\,\rm kpc$ and adopting an empirical relation between the equivalent width of Mg\,\textsc{ii} absorption and the column density of neutral hydrogen~\cite{Lan17}, we derive mass outflow rate $\dot{M}_{\rm out}$ and mass loading factor $\eta\equiv \dot{M}_{\rm out} / {\rm SFR}$, as shown in the bottom-right panels of Figure~\ref{fig:map}. 
However, an important caveat should be noted that $\dot{M}_{\rm out}$ and $\eta$ are secondary derived properties and are subject to substantial uncertainties which arise primarily from estimating the gas column densities (see Materials and Methods for further explanation).

As shown, $\rm EW_{out}$ increases notably with increasing $\rm SFR$, while shows relatively weak dependence on $M_*$.
We observe that the mean velocity of the outflowing gas is more correlated with $M_*$, whereas the dependence on $\rm SFR$ at fixed stellar mass is much weaker.
Similar dependence of $v_{\rm out}$ on $M_*$ for low-mass galaxies has also been found in the energy-driven wind in simulations~\cite{Mitchell20}, suggesting that the cool-phase gas traced by Mg\,\textsc{ii} is basically in an energy-conserving regime.
On the other hand, $\dot{M}_{\rm out}$ and $\eta$ have strong dependence on $\rm SFR$, {consistent with} the stellar feedback-driven outflow scenario~\cite{Rubin14}.
Our findings offer valuable observational insights into the effects of stellar feedback, providing constraints on the small-scale feedback mechanisms implemented in simulations.

The sample covers a broad redshift range, allowing us to investigate the above results across different redshift intervals.
The results are shown in Figure~\ref{fig:evolution}. 
The correlation between $v_{\rm out}$ and $M_*$ is nearly identical for the three redshift bins, which can be written as an empirical relation:  
\begin{equation}
\frac{v_{\rm out}}{\rm km\ s^{-1}} = (85.3\pm2.7) \cdot\log \left( 
 \frac{M_*}{10^{10}\,\rm M_\odot} \right) + (269.8\pm 0.9). \label{eq:vout}
\end{equation}
This shows that $v_{\rm out}$ increases $\sim85\, \rm km\,s^{-1}$ when the stellar mass increases 1 dex, with the typical mean outflow velocity to be $\sim 270\,\rm km\,s^{-1}$ for a galaxy with stellar mass of $10^{10} \,\rm M_{\odot}$.
Using an alternative spectra modelling method to extract outflow properties (as detailed in the Supplementary Materials), we also find $v_{\rm out}$ increases monotonically with $M_*$ regardless of redshift.
As displayed in figure~\ref{fig:scaling}, the relation derived from the alternative method yields $v_{\rm out}$ that are larger than those from the main method by $\sim 50\,\rm km\,s^{-1}$, with a slightly flatter slope.
The Mg\,\textsc{ii} absorption-line-derived outflow velocities taken from literature~\cite{Weiner09,Bordoloi14} are presented in panel A for comparison.
The relation between $v_{\rm out}$ and $\rm SFR$ is less tight across different redshift bins, as reflected by a slightly lower Pearson correlation coefficient (0.93 compared with 0.98), indicating that it may arise as a secondary effect of the underlying correlation between $v_{\rm out}$ and $M_*$.
Similarly, we find the relation between $\dot{M}_{\rm out}$ and $\rm SFR$ almost independent of redshift, which is
\begin{equation}
    \frac{\dot{M}_{\rm out}}{\rm M_\odot\, yr^{-1}} = (38.4 \pm 0.7) \cdot \log \left( \frac{\rm SFR}{\rm 10\, M_\odot\,yr^{-1}}\right) + (34.5\pm 0.3).\label{eq:mout}
\end{equation}
The alternative method, which produces larger $v_{\rm out}$ and smaller $\rm EW_{out}$ than the main method, yields a consistent relation.
On the other hand, $\dot{M}_{\rm out}$ as a function of $M_*$ for different redshifts shows substantial deviations, each exhibiting a nearly constant value. 
This again suggests that the outflow mass reflect the level of overall star formation, and the evolution of the outflow properties is primarily driven by the evolution of $\rm SFR$ along the cosmic time.

\subsection*{Discussion}

Although substantial uncertainties remain in both the mass outflow rate and the mass-loading factor, these quantities are nevertheless widely employed to constrain the underlying physical mechanisms of galactic feedback~\cite{Chisholm17, Swinbank19, Xu22, Avery22, Lyu26}.
While the hot phase of the CGM occupies a large fraction of the volume, the cool, neutral phase gas traced by Mg\,\textsc{ii} absorption may dominate the total mass budget of the outflowing gas~\cite{Banda-Barragan21}, rendering it of particular physical importance.
Current theoretical frameworks fundamentally bifurcate into two paradigms: (i) energy-driven flows where supernova (SNe) energy injection dominates radiative losses, and (ii) momentum-driven regimes where radiation pressure governs outflow dynamics~\cite{Veilleux20}.
In energy-driven scenario, the injected energy scale with $\rm SFR$~\cite{Leitherer99}, $\dot{E}\propto \rm SFR$, predicting a mass-loading factor scaling $\eta \propto v_{\rm out}^{-2}$. Conversely, momentum-driven models yield $\eta \propto v_{\rm out}^{-1}$~\cite{Murray05}. 
We therefore show the relation between $\eta$ and $v_{\rm out}$ in panel A of Figure~\ref{fig:eta}, excluding massive systems with $M_*>10^{10}\,\rm M_\odot$ where contamination from  active galactic nuclei (AGNs) could suppress measurements of $\eta$ (see figure~\ref{fig:agn} for evaluation of the contamination).
We note that for high-redshift ($1.2\leq z\leq 1.7$) sample $\eta$ is lower than other bins especially for massive galaxies, which we speculate to be relevant to the higher $\rm SFR$ at high-$z$.
We fit the data to models with slope of $-1$ and $-2$, and calculate the Bayesian evidence to compare the models.
The nested sampling analysis~\cite{Buchner23} yields a Bayes factor of $\log K=5.38$, strongly supporting the energy-driven scenario.
We emphasize that the uncertainty in estimating $\eta$ is potentially substantial but unavoidable at the present stage. 
Nevertheless, within this framework, our results support a scenario in which radiative losses are subdominant compared to mechanical energy injection in galactic feedback processes.
The correlation between $\eta$ and $M_*$ presented in panel B of Figure~\ref{fig:eta} exhibits a declining trend that is consistent with the results of numerical simulations~\cite{Nelson15,Nelson19,Angles17,Mitchell20}.
Values of $\eta$ derived from other previous observational studies~\cite{Schroetter19,Perrota23,ReichardtChu25} are also included, revealing substantial variations that largely reflect differences in sample selection and adopted methodologies (see Supplementary Materials for a detailed discussion). 

To determine whether the outflow can escape from the host galaxy, we calculate its velocity decay during propagation, adopting a Navarro-Frenk-White (NFW) DMH model~\cite{Navarro96} and an exponential stellar disk~\cite{Miyamoto75}.
Assuming the outflow is launched at $\rm 5\, kpc$ from the stellar disk with the observed $v_{\rm out}$, we examine how the velocity varies as its kinetic energy is converted into gravitational potential energy, as shown in the top panel of Figure~\ref{fig:gravity}.
We find that although the launch velocity of outflows in massive galaxies is higher, the gas cannot travel further than 100~kpc before falling back into the galaxy~\cite{Tollet19, Xu25}.
In contrast, outflows in galaxies with $M_*\lesssim 10^{10}\,\rm M_\odot$ are more likely to escape from the gravitational potential and enrich the surrounding environment.
This phenomenon is supported by studies of CGM metal absorption along the minor-axes of galaxies using background quasars~\cite{Chen24}, as shown in the bottom panel of Figure~\ref{fig:gravity} (see the stacked CGM spectra in figure~\ref{fig:cgm_spec}). 
As the stellar mass exceeds $10^{10}\,{\rm M_{\odot}}$, the EW(Mg\,\textsc{ii}) in the CGM between 20–150~kpc is substantially suppressed at the minor-axes, where the CGM is thought to be enriched by the bi-conical galactic outflows~\cite{Guo23}.
Given that the CGM Mg \textsc{ii} gas observed along the minor axes is likely associated with outflowing material from their host galaxies \cite{Nelson19}, the coincident turnover at $M_* \sim 10^{10}\,\mathrm{M_\odot}$ reinforces the down-the-barrel result that outflows are suppressed in massive DMHs.
Conversely, the cool outflows in low-mass galaxies can effectively escape from the gravitational potential.
This provides critical evidence that stellar feedback causes the baryon deficiency at the low-mass end of the galaxy population~\cite{McGaugh10}, emphasizing its power to  redistribute the baryonic material beyond the DMHs.

\subsection*{Materials and Methods}

\subsubsection*{Sample selection}

Focusing on star-forming galaxies, our sample is drawn from emission line galaxies (ELGs) in DESI EDR~\cite{DESIsample, DESIelg}. 
We implement a redshift cut of $z>0.3$ to ensure spectral coverage of the Mg\,\textsc{ii}\,$\lambda\lambda$\,2796,2803 doublet.
Considering the low S/N in the blue-tail of DESI spectra which is compromised by the low transmission in the blue spectrograph~\cite{DESIinst}, we further restrict the sample to $z>0.6$.
The stellar mass of each galaxy is estimated through stellar population synthesis fitting using Code Investigating GALaxy Emission (CIGALE)~\cite{Boquien19,Yang20,Yang22}, which utilizes the broad-band \textit{g}, \textit{r}, \textit{z}, \textit{W1} and \textit{W2} photometry from the DESI Legacy Imaging Surveys, and spectrophotometry of 10 artificial bands via convolution with the DESI optical spectra~\cite{Zou24}.
The star formation rates are estimated based on the flux of [O\,\textsc{II}]\,$\lambda\lambda$\,3726,3729 emission line~\cite{Schroetter19}, using the relation
\begin{equation}
    \frac{\rm SFR(O\,\textsc{ii})}{\rm M_\odot\, yr^{-1}} = 4.1\times 10^{-42}\, \frac{L(\rm O\,\textsc{ii})}{\rm erg\,s^{-1}},
\end{equation}
where the luminosity has been corrected for both Milky Way extinction~\cite{Schlegel98} and the intrinsic galaxy extinction~\cite{Garn10}, assuming the Chabrier initial mass function~\cite{Chabrier03} and Calzetti extinction law~\cite{Calzetti00}.

\begin{figure}
    \centering
    \includegraphics[width=0.9\linewidth]{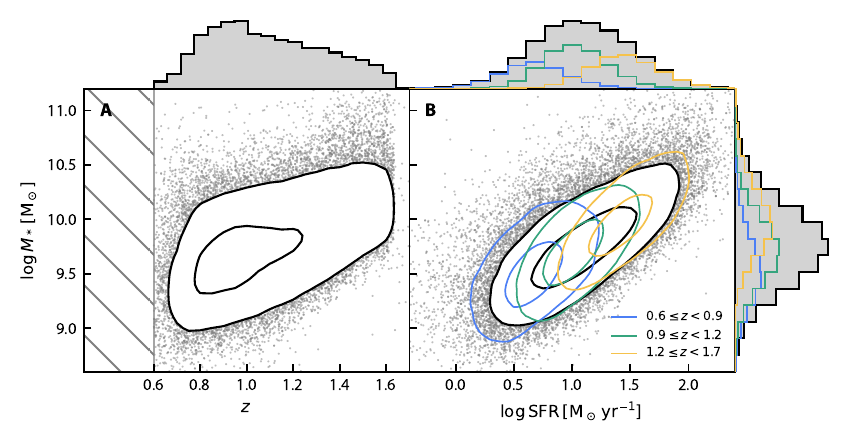}
    \caption{
    \textbf{Sample parameter distribution.}
    \textbf{(A)}
    $\log M_*$-$z$ distribution.
    The inner and outer black contours enclose 68\% and 95\% of the sample, respectively.
    For visual clarity, only 10\% of the points falling outside the outer contour are plotted.
    The histogram at the top of this panel shows the one-dimensional distribution of the sample redshifts.
    \textbf{(B)}
    $\log M_*$-$\log \rm SFR$ distribution.
    The blue, green and yellow contours indicate the distribution of samples with $0.6<z<0.9$, $0.9<z<1.2$ and $1.2<z<1.7$, respectively.
    Similar to the black contours, these enclose 68\% and 95\% of the corresponding redshift subsamples.
    The black histogram at the top of this panel displays the distribution of SFR for the samples, while the black histogram on the right shows the distribution of $M_*$.
    This histograms in the color of blue, green and yellow indicates the distribution of sample at different redshift bins.
    }
    \label{fig:sample}
\end{figure}

We apply the following selection criteria to the sample to exclude unphysical outliers. (i) The uncertainty in $M_*$ is less than 0.3 dex, ensuring reliable mass measurements; (ii) $10^6\,{\rm M_\odot}<M*<10^{14}\,{\rm M_\odot}$ and $10^{-4}\,{\rm M_\odot\,yr^{-1}}<{\rm SFR}<10^{4}\,{\rm M_\odot\,yr^{-1}}$.
After applying the cuts, our final sample comprises 394,909 ELGs spanning $0.6<z<1.7$, which is well representative of the general star-forming galaxies at these redshifts, with median stellar mass of $10^{9.8}\rm\, M_\odot$ and median SFR of $11.8 \rm\, M_\odot\,yr^{-1}$.
The distributions of $M_*$, $\rm SFR$ and $z$ are illustrated in Figure~\ref{fig:sample}.

\subsubsection*{Spectral Stacking and the Mg\,\textsc{ii} Profile}

We begin by shifting all individual spectra to rest-frame wavelengths determined by the spectroscopic redshifts, which were primarily inferred from strong emission line fitting~\cite{Guy2023}.
They are then resampled onto a common grid with a wavelength interval of $\Delta \lambda$ using \textsc{SpectRes}~\cite{Carnall17}.
Given that the DESI spectra have a native pixel size of 0.8~$\mathring{\rm A}$, we adopt $\Delta\lambda = 0.6~\mathring{\rm A}$ to avoid oversampling, taking into account the minimum redshift of our sample ($z_{\rm min}=0.6$).
Following this, a median filter with a smoothing window of 60\,$\rm \mathring{A}$ is applied before a fifth-order polynomial is fitted to the smoothed spectrum, deriving the continuum for the individual galaxy spectrum.
Throughout the aforementioned process, the prominent absorption and emission lines are masked: (i) $\pm 20$\,$\rm \mathring{A}$ around Mg\,\textsc{ii}\,$\lambda\lambda$2796,2803; (ii) $\pm 10\,{\rm \mathring{A}}$ around Mg\,\textsc{i}\,$\lambda$2851 and (iii) $\pm 40$\,$\rm \mathring{A}$ around Fe\,\textsc{ii}\,$\lambda$2600.
We then obtain the stacked spectra by calculating the median of the continuum-normalized spectra in different sample bins. 
Focusing on the outflow features in the Mg\,\textsc{ii}\,$\lambda\lambda$2796,2803 absorption profile, we have examined that most stacked spectra show flat continuum outside the absorption troughs.
However, at high mass end, a few stacked spectra exhibit continuum elevations around Mg\,\textsc{ii} absorptions up to $\sim 10\%$ above unity, extending out to $\sim 2000\,\rm km\,s^{-1}$.
This phenomenon may arise from the inclusion of AGNs producing broad Mg\,\textsc{ii} emission lines, which could affect the measurements of the outflow properties for the most massive galaxy bins.
We therefore perform another normalization to the stacked spectra to remove this large-scale residual, by dividing a triple-order polynomial fitting of the continuum while masking out 2785-2810\,$\rm \mathring{A}$~\cite{Zhu15}. 
For consistency, all stacked spectra are re-normalized.
Our test reveals that since the first normalizations to the individual spectra are reasonable enough, this second normalization to the composite spectra affects the derived $v_{\rm out}$ and $\rm EW_{out}$ to a very limited extent ($<6.8\%$).
The uncertainties in the stacked spectra are quantified through bootstrap resampling: the stacking procedure is repeated 1,000 times with replacement, with the 16–84th percentile range of the bootstrap realizations defining the $1\sigma$ confidence interval for each spectral bin.

The stacked spectra corresponding to the bins in Figure~\ref{fig:evolution} in the main text are presented in figure~\ref{fig:outflow_spec}. 
The stacked spectra exhibit an Mg\,\textsc{ii} absorption doublet ratio close to unity, in agreement with profiles reported in previous down-the-barrel observations~\cite{Erb12, Rubin14, Perrota23, Pessa24, Valentino25}.
Such a ratio is expected when the absorbing medium is optically thick and the absorption is saturated.
The intrinsically narrow, deep Mg\,\textsc{ii} absorption features are broadened and become shallower owing to instrumental line-spread effects, given the limited spectral resolution of DESI. 
In addition, the stacking procedure combines systems with diverse kinematic properties into a single composite spectrum, which further broadens the resultant absorption profile.
A potentially partial covering fraction of the Mg\,\textsc{ii}-bearing outflowing gas may also give rise to the shallow, yet saturated, absorption troughs.
As shown all stacked spectra display globally blue-shifted Mg\,\textsc{ii}\ absorption, which is a hallmark of galactic-scale outflows.
P-Cygni-like emission features are observed in a fraction of the composite spectra, and could be produced by the ISM of galaxies and/or outflows moving in the opposite direction by the means of resonant emission scattering~\cite{Martin13,Scarlata15,Carr18}.
The relative strength of the emission components is presumably correlated with the intensity of the ultra-violet (UV) continuum of the host galaxies, which naturally accounts for the decreasing contribution of Mg\,\textsc{ii} emission with increasing stellar mass.

Observations of Mg\,\textsc{ii} absorption on the circumgalactic scale using background quasars also yield doublet ratios close to 1~\cite{Lan18,Chen24}.
With the ionizing energy close to that of hydrogen, Mg\,\textsc{ii} traces cool, predominantly neutral gas with characteristic temperatures of around $10^4\,\rm K$.
This cool gas is detected preferentially along the minor axes of galaxies, which is generally interpreted as tracing the outflowing material~\cite{Schroetter19}.
Following the interaction between hot ionized outflows and the multiphase ISM, the cool, neutral gas traced by Mg\,\textsc{ii} is expected to form numerous dense clumps~\cite{Lan17,Banda-Barragan21}, further supporting our assumption that the observed Mg\,\textsc{ii} absorption is optically thick.

\subsubsection*{Outflow Properties}

\begin{figure}
    \centering
    \includegraphics[width=\textwidth]{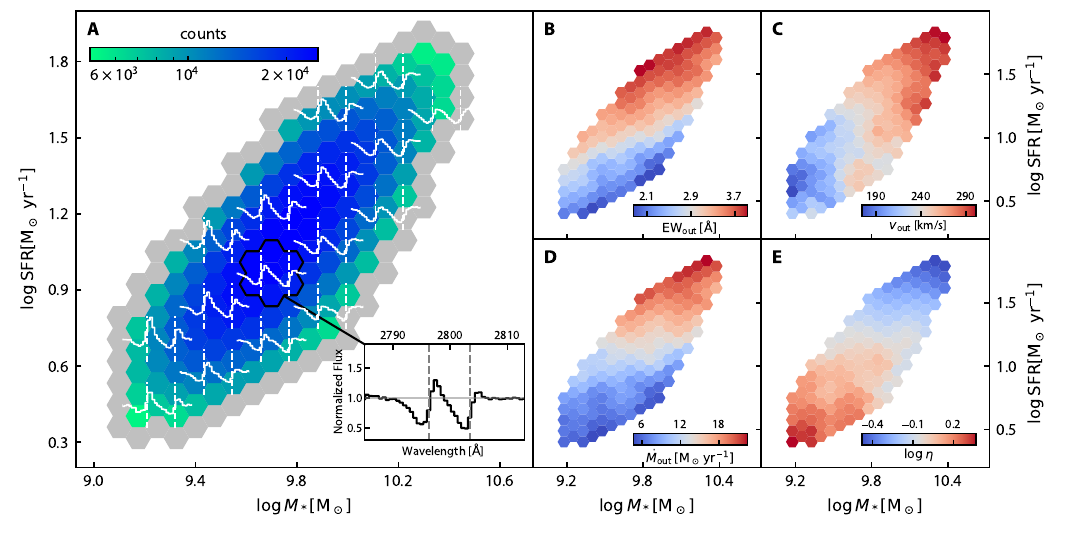}
    \caption{
    \textbf{Outflow properties of samples with $0.6<z<1.7$ on the $M_*$-$\rm SFR$ diagram.}
    \textbf{(A)} The distribution of samples, where colors of the hexagons indicate the total galaxy count per bin (including adjacent bins for stacking).
    Grey hexagons are excluded from color mapping due to insufficient neighboring bins, but the sample within them are used in the stacking of adjacent inner bins.
    Some of the stacked spectra are plotted over the hexagons with the two vertical dashed lines indicating the rest-frame wavelengths of Mg\,\textsc{ii}\,$\lambda$2796 and Mg\,\textsc{ii}\,$\lambda$2803, highlighting the prevalence of blueshifted outflow absorption features.
    The inset illustrates the stacked spectrum derived from galaxies within the black contour. 
    \textbf{(B)} The map of outflow equivalent width ($\rm EW_{out}$) derived from the stacked spectra.
    \textbf{(C)} The map of mean outflow velocity ($v_{\rm out}$) derived from the stacked spectra.
    \textbf{(D)} The map of mass outflow rate ($\dot{M}_{\rm out}$).
    \textbf{(E)} The map of mass-loading factor ($\eta\equiv \dot{M}_{\rm out}/\mathrm{SFR}$).
    }
    \label{fig:map}
\end{figure}

\begin{figure}
    \centering
    \includegraphics[width=\textwidth]{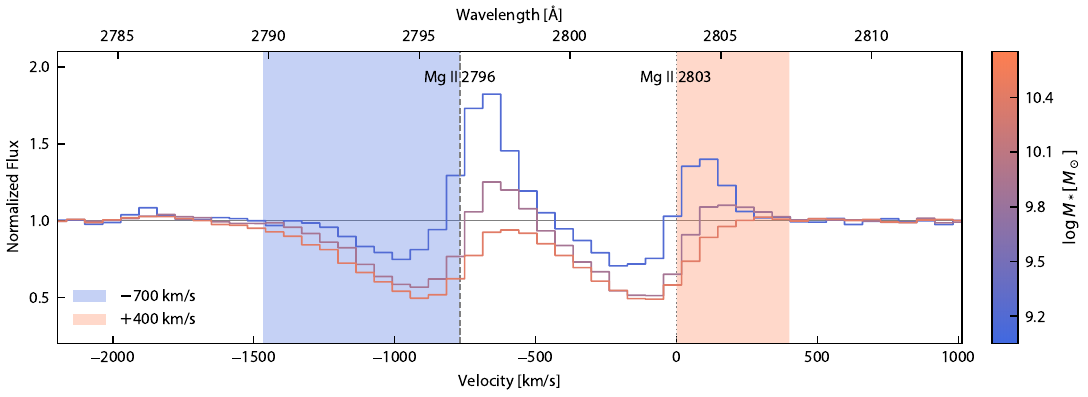}
    \caption{
    \textbf{Illustration of the \textsc{Boxcar} method for measuring outflow properties on the stacked spectra.}
    The color of the curves corresponds to the $M_*$ of the galaxies.
    The dashed and dotted lines indicate the rest-frame wavelengths of Mg\,\textsc{ii} $\lambda$2796 and Mg\,\textsc{ii} $\lambda$2803, respectively.
    All composite spectra presented here are obtained by stacking 20,000 individual galaxy spectra.
    The blue and orange shaded regions correspond to velocity range of $\rm -700$ to $\rm 0\,km\,s^{-1}$ relative to Mg\,\textsc{ii}\,$\lambda$2796 and $0$ to $\rm 400\,km\,s^{-1}$ relative to Mg\,\textsc{ii}\,$\lambda$2803, respectively.
    These two wavelength intervals are used to calculate outflow properties throughout this work. 
    Labels at the top of the figure indicate the rest-frame wavelengths, while those at the bottom show velocities relative to Mg\,\textsc{ii}\,$\lambda$2803.
    }
    \label{fig:boxcar}
\end{figure}

Although the blueshift in Mg\,\textsc{ii} absorption is apparent (e.g., in panel A of Figure \ref{fig:map}), the P-Cygni-like profile, which is found in most composite spectra, brings extra complexity in characterizing the outflows.
Considering the insufficient spectral resolution of the stacked spectra to resolve absorption of different optical depths, we struggle to make the simplest assumptions to make our results as model-independent as possible.
Based on the claim that the outflows are optically thick, we assume that the absorption profiles of Mg\,\textsc{ii}\,$\lambda$2796 and Mg\,\textsc{ii}\,$\lambda$2803 are identical.
Following the \textsc{Boxcar} method~\cite{Rubin10, Bordoloi14},
we exclude the spectral range between the Mg\,\textsc{ii} doublet, where the scattered emission and the blueshifted absorption are mixed altogether. 
By assuming saturation at $v=0~\rm km\,s^{-1}$, we merge the spectrum in the range of $-700$ to $0\rm~km\,s^{-1}$ relative to Mg\,\textsc{ii}\,$\lambda$2796 and $0$ to $400\rm~km\,s^{-1}$ relative to Mg\,\textsc{ii}\,$\lambda$2803, which encompass essentially all of the detectable signal in our high S/N normalized spectra, to derive a profile representative of Mg\,\textsc{ii}.
As shown in Figure~\ref{fig:boxcar}, we derive outflow EW through subtracting the non-outflowing symmetrical component (indicated by the orange band) from the total absorption (indicated by the blue band).
In this method we assume no inflow in the composite spectra, considering the extremely low covering fraction of inflows indicated by the scarcity of observed inflows in star forming galaxies~\cite{Rubin12, Weldon23, Coleman24}. 
Although \textsc{Boxcar} method overlooks the complicated radiative transfer processes in the outflows, it is nearly model-independent, which can directly extract the outflow features from the observed absorption profile.
This is favored in our statistical analysis of a large sample, highlighting the different outflow properties of various galaxy populations.
In addition to the \textsc{Boxcar} method, we also test with a more delicate approach to subtract the potential contribution from emission via model fitting; however, because this alternative method introduces additional uncertainty, we regard it as a complementary analysis (see Supplementary Materials for details). 

We calculate the $\rm EW$ of the blue-shifted absorption relative to Mg\,\textsc{ii}\,$\lambda$2796 and the red-shifted profile relative to Mg\,\textsc{ii}\,$\lambda$2803 as $\rm EW_{blue}$ and $\rm EW_{red}$, based on the two aforementioned wavelength intervals.
For simplicity, we decompose the absorption profile into two components: an outflowing component (out), characterized by a blue-shifted line center, and an interstellar component (sym), with a symmetric profile centered at the rest-frame wavelength.
Ignoring the mixed absorption and emission between 2796.35~$\rm \mathring{A}$ and 2803.53~$\rm \mathring{A}$, the total equivalent width of the doublet defined as $\rm EW_{tot}\equiv EW_{out} + EW_{sym}$ and the equivalent width of the outflowing component $\rm EW_{out}$ are estimated as
\begin{align} 
    \rm EW_{tot} &= 2 (\rm EW_{blue} + EW_{red}),\\
    \rm EW_{out} &= 2 (\rm EW_{blue} - EW_{red}),\label{eq:out}
\end{align} 
where the coefficient of 2 converts from EW of the single merged profile to that of the doublets.
figure~\ref{fig:ew} presents the measured $\rm EW_{out}$ as functions of $M_*$ and $\rm SFR$, together with results from previous Mg\,\textsc{ii} absorption stacking studies~\cite{Weiner09,Rubin10,Bordoloi14} for comparison.
We note that older studies observed a weak positive correlation between $\rm EW_{out}$ and $\log M_*$, whereas our data reveals that $\rm EW_{out}$ remains almost constant in a given redshift bin.
On the other hand we observe positive $\rm EW_{out}$-$\log \rm SFR$ correlation in each redshift bin, which is consistent with those studies.
This implicates that the $\rm EW_{out}$-$\log M_*$ relation is a biproduct of the strong $M_*$-$z$ relation, strengthening that outflows vary with star-formation rates within galaxies which evolves through the cosmic time.

The EW-weighted mean velocity of the merged profile can be calculated as $v_{\rm tot}$.
Considering that the velocity of the symmetric component is 0, the velocity of the outflowing component is calculated as
\begin{equation}
    v_{\rm out} = \frac{\rm EW_{tot}}{\rm EW_{out}} \cdot v_{\rm tot}.
\end{equation}

We then adopt an empirical relation of CGM Mg\,\textsc{ii} absorbers to convert EW(Mg\,\textsc{ii}) into the column density of neutral hydrogen~\cite{Lan17}, as

\begin{equation}
    N_{\rm HI} = A\left( \frac{\rm EW_{\lambda 2796}}{1\,\rm \mathring{A}} \right)^\alpha (1+z)^\beta \label{eq:nhi}
\end{equation}
where $\alpha=1.69\pm0.13$, $\beta=1.88\pm 0.29$ and $A=10^{18.96\pm 0.10}\,\rm cm^{-2}$.
The $\rm EW_{\lambda 2796}$ is the equivalent width of Mg\,\textsc{ii}\,$\lambda$2796 absorption,
which is half of the total doublet EW according to the aforementioned assumptions.
The mass outflow rate is then calculated assuming the geometry of the outflow to be an expanding thin spherical shell~\cite{Weiner09}:
\begin{equation} \label{eq:moutcalc}
    \dot{M}_{\rm out} = 22\,\mathrm{M_\odot\,yr^{-1}} \cdot \frac{N_{\rm HI}}{10^{20}\,\rm cm^{-2}} \cdot \frac{r}{r_{\rm launch}}\cdot \frac{v_{\rm out}}{300\,\mathrm{km\,s^{-1}}},
\end{equation}
where the covering fraction of the outflow is assumed to be 1, as a composite spectrum effectively averages over all viewing angles and the composite integrates over covered and uncovered lines of sight and clumpiness within the outflows~\cite{Weiner09}. 
If a non-unity covering fraction of 0.5 is assumed, the derived $\dot{M}_{\rm out}$ and $\eta$ would shift up $\sim 0.2\,\rm dex$ (see Supplementary Materials for discussion).
The launch radius $r_{\rm launch}$ is expected to be of the order of the galaxy radius, which is poorly constrained for our $z>0.6$ sample because of the low spatial resolution of DESI legacy imaging surveys.
Here we assume the launch radius of the outflow is fixed at $r_{\rm launch}=5\rm\,kpc$, which is the typical value adopted in the literature~\cite{Burchett21,Zabl21,Perrota23}, as an order-of-magnitude estimate.
From this equation, $\dot{M}_{\rm out}$ is proportional to the product of $v_{\rm out}$ and a power of $\rm EW_{out}$.

The uncertainties in $\rm EW_{out}$ and $v_{\rm out}$ are quantified through an analysis of 1,000 stacked spectra generated via bootstrap resamplings. 
We apply the \textsc{Boxcar} method on these resampled spectra and determine the 1$\sigma$ confidence intervals for $\rm EW_{out}$ and $v_{\rm out}$ by calculating the 16-84th percentile ranges of the bootstrap distributions.
The uncertainties in $v_{\rm out}$, $\rm EW_{\rm out}$ and parameters in the empirical relations (Equation~\ref{eq:nhi}) are propagated into the uncertainty in $\dot{M}_{\rm out}$.
The uncertainty in $\eta$ is further derived using error propagation, assuming that the uncertainties in the SFR measurements are negligible, since the SFR in a given bin spans a relatively small range compared with the estimated uncertainty in $\dot{M}_{\rm out}$.

We emphasize that the derivation of $\dot{M}_{\rm out}$ and $\eta$ is subject to potentially large uncertainties, primarily due to the intrinsic degeneracy of equivalent width with column density, Doppler parameter, velocity dispersion and covering fraction. Although the uncertainty in the empirical conversion relation partially encapsulates these effects, the inferred values are intended as secondary, illustrative estimates.

\subsubsection*{Gravititional Potential Modelling}

We determine the mass of DMHs corresponding to different stellar mass through the $M_*$-$M_{\rm halo}$ relation from Moster et al.~\cite{Moster2010}, and assume an NFW density profile~\cite{Navarro96} which is expressed as
\begin{equation}
    \frac{\rho(r)}{\rho_{\rm crit}} = \frac{\delta_c}{(r/r_s)(1+r/r_s)^2},
\end{equation}
where $r_s=r_{200}/c$ is a characteristic radius and $\rho_{\rm crit}=3H^2/8\pi G$.
$\delta_c$ is a dimensionless parameter defined as 
\begin{equation}
    \delta_c = \frac{200}{3}\frac{c^3}{[\ln (1+c)-c/(1+c)]},
\end{equation}
where $c$ is the concentration parameter depending on $M_{\rm halo}$, which is taken from Equation 8 from Maccio et al. 2007~\cite{Maccio07}.
By integrating the density profile, we obtain the gravity potential of an NFW halo as
\begin{equation}
\Phi_{\rm NFW}(r) = - 4\pi G \rho_{\rm crit} \delta_c r_s^3 \frac{1}{r+r_s}.
\end{equation}
Additionally, we consider the potential of an exponential disk~\cite{Miyamoto75} in the form of 
\begin{equation}
\Phi_{\rm disk}(R,z)=\frac{GM}{\{R^2 + [a + (z^2+b^2)^{1/2}]^2\}^{1/2}},
\end{equation}
where we assume $a=3\rm\, kpc$, $b=0.3\rm\, kpc$ and $R=0$, considering the physical size of our sample is comparable to 3\,kpc.
The total gravity potential is obtained through summing $\Phi_{\rm NFW}$ and $\Phi_{\rm disk}$.

\begin{figure}
    \centering
    \includegraphics[width=\textwidth]{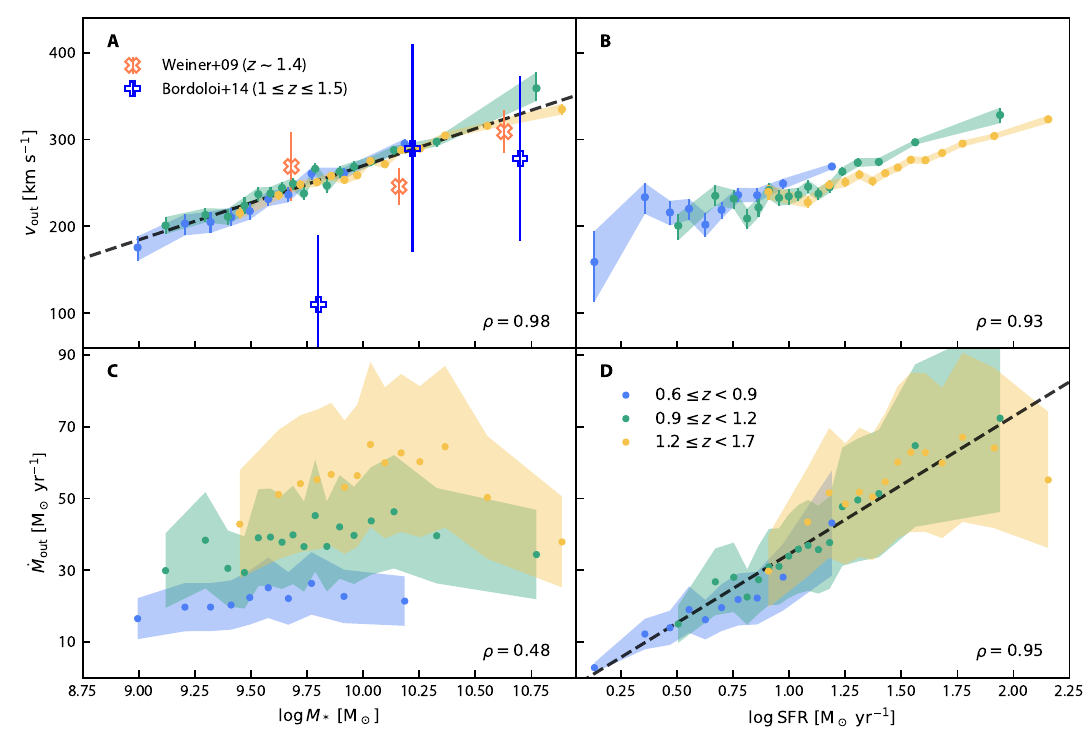}
    \caption{
    \textbf{
    Scaling relations between outflow properties and galaxy properties at different redshifts.
    }
    \textbf{(A)}
    $v_{\rm out}$ as a function of $\log M_*$.
    The samples are divided into three bins by redshift.
    The blue, green and yellow points represent the samples with $0.6\leq z<0.9$, $0.9\leq z<1.2$ and $1.2\leq z < 1.7$, respectively.
    The open symbols denote Mg\,\textsc{ii} outflow velocity measurements~\cite{Weiner09, Bordoloi14} from the literature.
    The value of each point is derived with a composite spectrum of 10,000 spectra, with the shaded area indicating 1$\sigma$ uncertainties from bootstrapping analysis.
    The dashed line is the best-fit relation, see Equation~\ref{eq:vout}.
    The Pearson correlation coefficient computed using data points across all redshift bins is displayed at the bottom-right corner of the panel.
    \textbf{(B)}
    $v_{\rm out}$ as a function of $\log \rm SFR$.
    \textbf{(C)}
    $\dot{M}_{\rm out}$ as a function of $\log M_*$.
    The shaded area indicates 1$\sigma$ uncertainties propagated from the uncertainty in $\rm EW_{out}$ derived from bootstrap sampling and the uncertainty in estimating H\,\textsc{i} column density from $\rm EW_{out}$.
    \textbf{(D)} 
    $\dot{M}_{\rm out}$ as a function of $\log \rm SFR$.
    The dashed line is the best-fit relation, see Equation~\ref{eq:mout}.
    }
    \label{fig:evolution}
\end{figure}

\begin{figure}
    \centering
    \includegraphics[width=\textwidth]{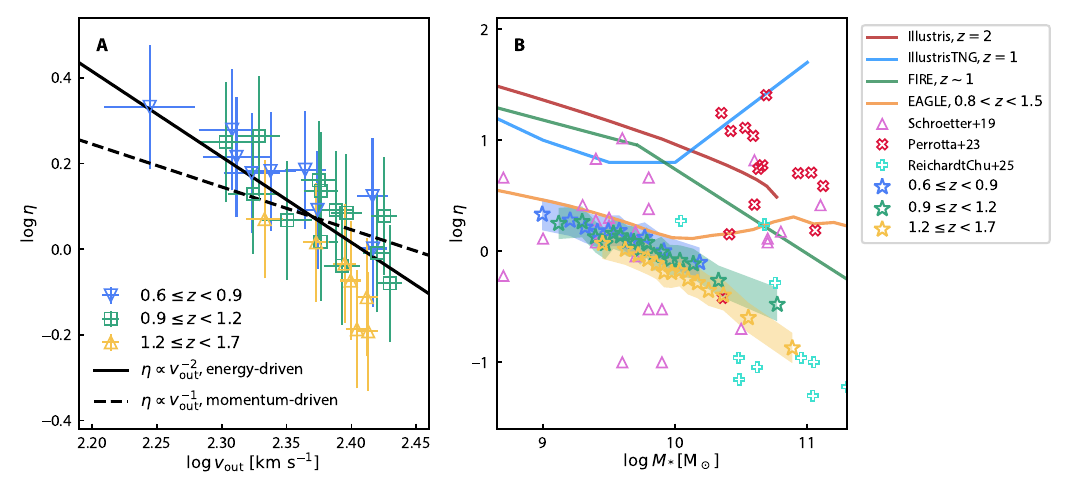}
    \caption{
    \textbf{The mass-loading factor.}
    \textbf{(A)}
    The mass-loading factor as a function of outflow velocity, distinguishing energy-driven and momentum-driven outflow regimes.
    The data points colored by redshift are taken from the left panels of Figure~\ref{fig:evolution}. 
    To mitigate potential contamination from AGN activity in massive galaxies, we exclude data points with $M_*>10^{10}\,\rm M_\odot$.
    The solid and dashed lines show best-fit models assuming fixed power-law exponents of $-2$ (energy-driven scenario) and $-1$ (momentum-driven scenario), respectively.
    \textbf{(B)}
    The derived mass-loading factor as a function of stellar mass in comparison with simulations and other observations.
    The open stars and shaded area represent $\eta$ and the 1$\sigma$ uncertainties derived in this work, with blue, green and yellow indicating $0.6\leq z<0.9$, $0.9\leq z<1.2$ and $1.2\leq z < 1.7$, respectively.
    The lines are taken from papers of corresponding simulations, the $M_*$ values of which are converted from $M_{\rm halo}$ through $M_*$-$M_{\rm halo}$ relation~\cite{Moster13}.
    The other open symbols represent results from observations.
    }
    \label{fig:eta}
\end{figure}

\begin{figure}
    \centering
    \includegraphics[width=0.6\textwidth]{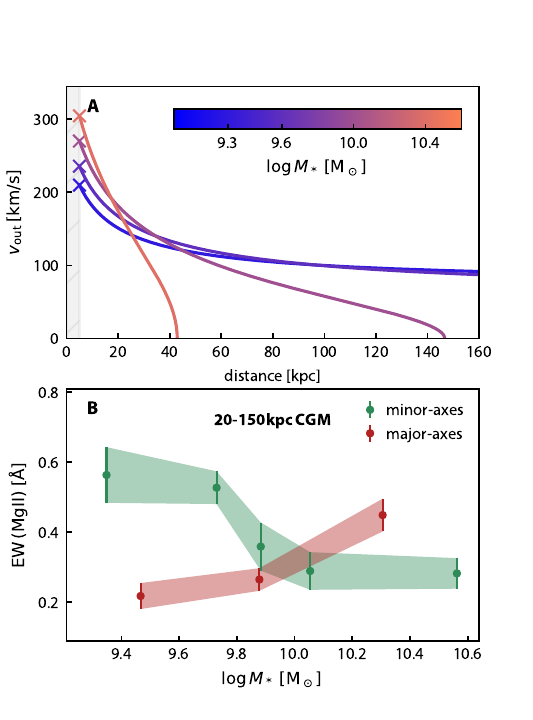}
    \caption{
    \textbf{
    The alignment of hosting gravitational bound outflows and minor-axes CGM suppresion at $M_*\sim 10^{10}\,\rm M_\odot$.
    }
    \textbf{(A)}
    The velocity decay of outflows.
    The diagonal crosses represent the observed outflow velocities, which are obtained through Equation~\ref{eq:vout} according to the given $M_*$ of galaxies.
    The four curves with increasing launching velocity represent velocity decay of outflows in galaxies with stellar masses of $10^{9.3}\,\rm  M_\odot$, $10^{9.6}\,\rm  M_\odot$, $10^{10.0}\,\rm  M_\odot$ and $10^{10.4}\,\rm  M_\odot$.
    \textbf{(B)}
    Total Mg\,\textsc{ii} absorption equivalent width in the 20-150\,kpc CGM surrounding galaxies derived using background quasars.
    The green (red) symbols represents CGM at the minor (major)-axes of the galaxy. 
    }
    \label{fig:gravity}
\end{figure}

\clearpage 

%

\begin{thebibliography}{10}
\providecommand{\url}[1]{\texttt{#1}}
\expandafter\ifx\csname urlstyle\endcsname\relax
  \providecommand{\doi}[1]{doi:\discretionary{}{}{}#1}\else
  \providecommand{\doi}{doi:\discretionary{}{}{}\begingroup \urlstyle{rm}\Url}\fi

\bibitem{vandeVoort11}
F.~{van de Voort}, J.~{Schaye}, C.~M. {Booth}, M.~R. {Haas}, C.~{Dalla Vecchia}, {The rates and modes of gas accretion on to galaxies and their gaseous haloes}. \emph{\mnras} \textbf{414}~(3), 2458--2478 (2011), \doi{10.1111/j.1365-2966.2011.18565.x}.

\bibitem{Thompson16}
T.~A. {Thompson}, E.~{Quataert}, D.~{Zhang}, D.~H. {Weinberg}, {An origin for multiphase gas in galactic winds and haloes}. \emph{\mnras} \textbf{455}~(2), 1830--1844 (2016), \doi{10.1093/mnras/stv2428}.

\bibitem{Veilleux05}
S.~{Veilleux}, G.~{Cecil}, J.~{Bland-Hawthorn}, {Galactic Winds}. \emph{\araa} \textbf{43}~(1), 769--826 (2005), \doi{10.1146/annurev.astro.43.072103.150610}.

\bibitem{Veilleux20}
S.~{Veilleux}, R.~{Maiolino}, A.~D. {Bolatto}, S.~{Aalto}, {Cool outflows in galaxies and their implications}. \emph{\aapr} \textbf{28}~(1), 2 (2020), \doi{10.1007/s00159-019-0121-9}.

\bibitem{Thompson24}
T.~A. {Thompson}, T.~M. {Heckman}, {Theory and Observation of Winds from Star-Forming Galaxies}. \emph{\araa} \textbf{62}~(1), 529--591 (2024), \doi{10.1146/annurev-astro-041224-011924}.

\bibitem{Behroozi10}
P.~S. {Behroozi}, C.~{Conroy}, R.~H. {Wechsler}, {A Comprehensive Analysis of Uncertainties Affecting the Stellar Mass-Halo Mass Relation for 0 < z < 4}. \emph{\apj} \textbf{717}~(1), 379--403 (2010), \doi{10.1088/0004-637X/717/1/379}.

\bibitem{Moster13}
B.~P. {Moster}, T.~{Naab}, S.~D.~M. {White}, {Galactic star formation and accretion histories from matching galaxies to dark matter haloes}. \emph{\mnras} \textbf{428}~(4), 3121--3138 (2013), \doi{10.1093/mnras/sts261}.

\bibitem{Puchwein13}
E.~{Puchwein}, V.~{Springel}, {Shaping the galaxy stellar mass function with supernova- and AGN-driven winds}. \emph{\mnras} \textbf{428}~(4), 2966--2979 (2013), \doi{10.1093/mnras/sts243}.

\bibitem{Vogelsberger20}
M.~{Vogelsberger}, F.~{Marinacci}, P.~{Torrey}, E.~{Puchwein}, {Cosmological simulations of galaxy formation}. \emph{Nature Reviews Physics} \textbf{2}~(1), 42--66 (2020), \doi{10.1038/s42254-019-0127-2}.

\bibitem{Hopkins14}
P.~F. {Hopkins}, \emph{et~al.}, {Galaxies on FIRE (Feedback In Realistic Environments): stellar feedback explains cosmologically inefficient star formation}. \emph{\mnras} \textbf{445}~(1), 581--603 (2014), \doi{10.1093/mnras/stu1738}.

\bibitem{Hopkins18}
P.~F. {Hopkins}, \emph{et~al.}, {FIRE-2 simulations: physics versus numerics in galaxy formation}. \emph{\mnras} \textbf{480}~(1), 800--863 (2018), \doi{10.1093/mnras/sty1690}.

\bibitem{Schaye15}
J.~{Schaye}, \emph{et~al.}, {The EAGLE project: simulating the evolution and assembly of galaxies and their environments}. \emph{\mnras} \textbf{446}~(1), 521--554 (2015), \doi{10.1093/mnras/stu2058}.

\bibitem{Vogelsberger14}
M.~{Vogelsberger}, \emph{et~al.}, {Introducing the Illustris Project: simulating the coevolution of dark and visible matter in the Universe}. \emph{\mnras} \textbf{444}~(2), 1518--1547 (2014), \doi{10.1093/mnras/stu1536}.

\bibitem{Nelson15}
D.~{Nelson}, \emph{et~al.}, {The illustris simulation: Public data release}. \emph{Astronomy and Computing} \textbf{13}, 12--37 (2015), \doi{10.1016/j.ascom.2015.09.003}.

\bibitem{DESIedr}
{DESI Collaboration}, \emph{et~al.}, {The Early Data Release of the Dark Energy Spectroscopic Instrument}. \emph{\aj} \textbf{168}~(2), 58 (2024), \doi{10.3847/1538-3881/ad3217}.

\bibitem{Brinchmann04}
J.~{Brinchmann}, \emph{et~al.}, {The physical properties of star-forming galaxies in the low-redshift Universe}. \emph{\mnras} \textbf{351}~(4), 1151--1179 (2004), \doi{10.1111/j.1365-2966.2004.07881.x}.

\bibitem{Noeske07}
K.~G. {Noeske}, \emph{et~al.}, {Star Formation in AEGIS Field Galaxies since z=1.1: Staged Galaxy Formation and a Model of Mass-dependent Gas Exhaustion}. \emph{\apjl} \textbf{660}~(1), L47--L50 (2007), \doi{10.1086/517927}.

\bibitem{Speagle14}
J.~S. {Speagle}, C.~L. {Steinhardt}, P.~L. {Capak}, J.~D. {Silverman}, {A Highly Consistent Framework for the Evolution of the Star-Forming ``Main Sequence'' from z \raisebox{-0.5ex}\textasciitilde 0-6}. \emph{\apjs} \textbf{214}~(2), 15 (2014), \doi{10.1088/0067-0049/214/2/15}.

\bibitem{Morton03}
D.~C. {Morton}, {Atomic Data for Resonance Absorption Lines. III. Wavelengths Longward of the Lyman Limit for the Elements Hydrogen to Gallium}. \emph{\apjs} \textbf{149}~(1), 205--238 (2003), \doi{10.1086/377639}.

\bibitem{Erb12}
D.~K. {Erb}, A.~M. {Quider}, A.~L. {Henry}, C.~L. {Martin}, {Galactic Outflows in Absorption and Emission: Near-ultraviolet Spectroscopy of Galaxies at 1 < z < 2}. \emph{\apj} \textbf{759}~(1), 26 (2012), \doi{10.1088/0004-637X/759/1/26}.

\bibitem{Rubin14}
K.~H.~R. {Rubin}, \emph{et~al.}, {Evidence for Ubiquitous Collimated Galactic-scale Outflows along the Star-forming Sequence at z \raisebox{-0.5ex}\textasciitilde 0.5}. \emph{\apj} \textbf{794}~(2), 156 (2014), \doi{10.1088/0004-637X/794/2/156}.

\bibitem{Perrota23}
S.~{Perrotta}, \emph{et~al.}, {Kinematics, Structure, and Mass Outflow Rates of Extreme Starburst Galactic Outflows}. \emph{\apj} \textbf{949}~(1), 9 (2023), \doi{10.3847/1538-4357/acc660}.

\bibitem{Pessa24}
I.~{Pessa}, \emph{et~al.}, {A galactic outflow traced by its extended Mg II emission out to a {\ensuremath{\sim}}30 kpc radius in the Hubble Ultra Deep Field with MUSE}. \emph{\aap} \textbf{691}, A5 (2024), \doi{10.1051/0004-6361/202450547}.

\bibitem{Valentino25}
F.~{Valentino}, \emph{et~al.}, {Gas outflows in two recently quenched galaxies at z = 4 and 7}. \emph{\aap} \textbf{699}, A358 (2025), \doi{10.1051/0004-6361/202553908}.

\bibitem{Rubin10}
K.~H.~R. {Rubin}, \emph{et~al.}, {The Persistence of Cool Galactic Winds in High Stellar Mass Galaxies between z \raisebox{-0.5ex}\textasciitilde 1.4 and \raisebox{-0.5ex}\textasciitilde1}. \emph{\apj} \textbf{719}~(2), 1503--1525 (2010), \doi{10.1088/0004-637X/719/2/1503}.

\bibitem{Bordoloi14}
R.~{Bordoloi}, \emph{et~al.}, {The Dependence of Galactic Outflows on the Properties and Orientation of zCOSMOS Galaxies at z \raisebox{-0.5ex}\textasciitilde 1}. \emph{\apj} \textbf{794}~(2), 130 (2014), \doi{10.1088/0004-637X/794/2/130}.

\bibitem{Lan17}
T.-W. {Lan}, M.~{Fukugita}, {Mg II Absorbers: Metallicity Evolution and Cloud Morphology}. \emph{\apj} \textbf{850}~(2), 156 (2017), \doi{10.3847/1538-4357/aa93eb}.

\bibitem{Mitchell20}
P.~D. {Mitchell}, J.~{Schaye}, R.~G. {Bower}, R.~A. {Crain}, {Galactic outflow rates in the EAGLE simulations}. \emph{\mnras} \textbf{494}~(3), 3971--3997 (2020), \doi{10.1093/mnras/staa938}.

\bibitem{Weiner09}
B.~J. {Weiner}, \emph{et~al.}, {Ubiquitous Outflows in DEEP2 Spectra of Star-Forming Galaxies at z = 1.4}. \emph{\apj} \textbf{692}~(1), 187--211 (2009), \doi{10.1088/0004-637X/692/1/187}.

\bibitem{Chisholm17}
J.~{Chisholm}, C.~A. {Tremonti}, C.~{Leitherer}, Y.~{Chen}, {The mass and momentum outflow rates of photoionized galactic outflows}. \emph{\mnras} \textbf{469}~(4), 4831--4849 (2017), \doi{10.1093/mnras/stx1164}.

\bibitem{Swinbank19}
A.~M. {Swinbank}, \emph{et~al.}, {The energetics of starburst-driven outflows at z {\ensuremath{\sim}} 1 from KMOS}. \emph{\mnras} \textbf{487}~(1), 381--393 (2019), \doi{10.1093/mnras/stz1275}.

\bibitem{Xu22}
X.~{Xu}, \emph{et~al.}, {CLASSY III. The Properties of Starburst-driven Warm Ionized Outflows}. \emph{\apj} \textbf{933}~(2), 222 (2022), \doi{10.3847/1538-4357/ac6d56}.

\bibitem{Avery22}
C.~R. {Avery}, \emph{et~al.}, {Cool outflows in MaNGA: a systematic study and comparison to the warm phase}. \emph{\mnras} \textbf{511}~(3), 4223--4237 (2022), \doi{10.1093/mnras/stac190}.

\bibitem{Lyu26}
C.~{Lyu}, \emph{et~al.}, {First Statistical Detection of Cool Gas Outflows with JWST Towards Cosmic Dawn}. \emph{arXiv e-prints} arXiv:2512.05622 (2025), \doi{10.48550/arXiv.2512.05622}.

\bibitem{Banda-Barragan21}
W.~E. {Banda-Barrag{\'a}n}, \emph{et~al.}, {Shock-multicloud interactions in galactic outflows - II. Radiative fractal clouds and cold gas thermodynamics}. \emph{\mnras} \textbf{506}~(4), 5658--5680 (2021), \doi{10.1093/mnras/stab1884}.

\bibitem{Leitherer99}
C.~{Leitherer}, \emph{et~al.}, {Starburst99: Synthesis Models for Galaxies with Active Star Formation}. \emph{\apjs} \textbf{123}~(1), 3--40 (1999), \doi{10.1086/313233}.

\bibitem{Murray05}
N.~{Murray}, E.~{Quataert}, T.~A. {Thompson}, {On the Maximum Luminosity of Galaxies and Their Central Black Holes: Feedback from Momentum-driven Winds}. \emph{\apj} \textbf{618}~(2), 569--585 (2005), \doi{10.1086/426067}.

\bibitem{Buchner23}
J.~{Buchner}, {Nested Sampling Methods}. \emph{Statistics Surveys} \textbf{17}, 169--215 (2023), \doi{10.1214/23-SS144}.

\bibitem{Nelson19}
D.~{Nelson}, \emph{et~al.}, {First results from the TNG50 simulation: galactic outflows driven by supernovae and black hole feedback}. \emph{\mnras} \textbf{490}~(3), 3234--3261 (2019), \doi{10.1093/mnras/stz2306}.

\bibitem{Angles17}
D.~{Angl{\'e}s-Alc{\'a}zar}, \emph{et~al.}, {The cosmic baryon cycle and galaxy mass assembly in the FIRE simulations}. \emph{\mnras} \textbf{470}~(4), 4698--4719 (2017), \doi{10.1093/mnras/stx1517}.

\bibitem{Schroetter19}
I.~{Schroetter}, \emph{et~al.}, {MusE GAs FLOw and Wind (MEGAFLOW) - III. Galactic wind properties using background quasars}. \emph{\mnras} \textbf{490}~(3), 4368--4381 (2019), \doi{10.1093/mnras/stz2822}.

\bibitem{ReichardtChu25}
B.~{Reichardt Chu}, \emph{et~al.}, {DUVET: sub-kiloparsec resolved star formation driven outflows in a sample of local starbursting disc galaxies}. \emph{\mnras} \textbf{536}~(2), 1799--1821 (2025), \doi{10.1093/mnras/stae2705}.

\bibitem{Navarro96}
J.~F. {Navarro}, C.~S. {Frenk}, S.~D.~M. {White}, {The Structure of Cold Dark Matter Halos}. \emph{\apj} \textbf{462}, 563 (1996), \doi{10.1086/177173}.

\bibitem{Miyamoto75}
M.~{Miyamoto}, R.~{Nagai}, {Three-dimensional models for the distribution of mass in galaxies.} \emph{\pasj} \textbf{27}, 533--543 (1975).

\bibitem{Tollet19}
{\'E}.~{Tollet}, A.~{Cattaneo}, A.~V. {Macci{\`o}}, A.~A. {Dutton}, X.~{Kang}, {NIHAO XIX: how supernova feedback shapes the galaxy baryon cycle}. \emph{\mnras} \textbf{485}~(2), 2511--2531 (2019), \doi{10.1093/mnras/stz545}.

\bibitem{Xu25}
Y.~{Xu}, \emph{et~al.}, {Stellar- and AGN-driven Outflows in JWST Galaxies at z = 3{\textendash}9: More Frequent, Wider Opening Angles, and Mostly Bounded}. \emph{\apj} \textbf{984}~(2), 182 (2025), \doi{10.3847/1538-4357/adc733}.

\bibitem{Chen24}
Z.~{Chen}, \emph{et~al.}, {The Circumgalactic Medium Traced by Mg II Absorption with DESI: Dependence on Galaxy Stellar Mass, Star Formation Rate, and Azimuthal Angle}. \emph{\apj} \textbf{981}~(1), 81 (2025), \doi{10.3847/1538-4357/ada942}.

\bibitem{Guo23}
Y.~{Guo}, \emph{et~al.}, {Bipolar outflows out to 10 kpc for massive galaxies at redshift z {\ensuremath{\approx}} 1}. \emph{\nat} \textbf{624}~(7990), 53--56 (2023), \doi{10.1038/s41586-023-06718-w}.

\bibitem{McGaugh10}
S.~S. {McGaugh}, J.~M. {Schombert}, W.~J.~G. {de Blok}, M.~J. {Zagursky}, {The Baryon Content of Cosmic Structures}. \emph{\apjl} \textbf{708}~(1), L14--L17 (2010), \doi{10.1088/2041-8205/708/1/L14}.

\bibitem{DESIsample}
{DESI Collaboration}, \emph{et~al.}, {DESI 2024 II: Sample Definitions, Characteristics, and Two-point Clustering Statistics}. \emph{arXiv e-prints} arXiv:2411.12020 (2024), \doi{10.48550/arXiv.2411.12020}.

\bibitem{DESIelg}
A.~{Raichoor}, \emph{et~al.}, {Target Selection and Validation of DESI Emission Line Galaxies}. \emph{\aj} \textbf{165}~(3), 126 (2023), \doi{10.3847/1538-3881/acb213}.

\bibitem{DESIinst}
{DESI Collaboration}, \emph{et~al.}, {Overview of the Instrumentation for the Dark Energy Spectroscopic Instrument}. \emph{\aj} \textbf{164}~(5), 207 (2022), \doi{10.3847/1538-3881/ac882b}.

\bibitem{Boquien19}
M.~{Boquien}, \emph{et~al.}, {CIGALE: a python Code Investigating GALaxy Emission}. \emph{\aap} \textbf{622}, A103 (2019), \doi{10.1051/0004-6361/201834156}.

\bibitem{Yang20}
G.~{Yang}, \emph{et~al.}, {X-CIGALE: Fitting AGN/galaxy SEDs from X-ray to infrared}. \emph{\mnras} \textbf{491}~(1), 740--757 (2020), \doi{10.1093/mnras/stz3001}.

\bibitem{Yang22}
G.~{Yang}, \emph{et~al.}, {Fitting AGN/Galaxy X-Ray-to-radio SEDs with CIGALE and Improvement of the Code}. \emph{\apj} \textbf{927}~(2), 192 (2022), \doi{10.3847/1538-4357/ac4971}.

\bibitem{Zou24}
H.~{Zou}, \emph{et~al.}, {A Large Sample of Extremely Metal-poor Galaxies at z < 1 Identified from the DESI Early Data}. \emph{\apj} \textbf{961}~(2), 173 (2024), \doi{10.3847/1538-4357/ad1409}.

\bibitem{Schlegel98}
D.~J. {Schlegel}, D.~P. {Finkbeiner}, M.~{Davis}, {Maps of Dust Infrared Emission for Use in Estimation of Reddening and Cosmic Microwave Background Radiation Foregrounds}. \emph{\apj} \textbf{500}~(2), 525--553 (1998), \doi{10.1086/305772}.

\bibitem{Garn10}
T.~{Garn}, P.~N. {Best}, {Predicting dust extinction from the stellar mass of a galaxy}. \emph{\mnras} \textbf{409}~(1), 421--432 (2010), \doi{10.1111/j.1365-2966.2010.17321.x}.

\bibitem{Chabrier03}
G.~{Chabrier}, {Galactic Stellar and Substellar Initial Mass Function}. \emph{\pasp} \textbf{115}~(809), 763--795 (2003), \doi{10.1086/376392}.

\bibitem{Calzetti00}
D.~{Calzetti}, \emph{et~al.}, {The Dust Content and Opacity of Actively Star-forming Galaxies}. \emph{\apj} \textbf{533}~(2), 682--695 (2000), \doi{10.1086/308692}.

\bibitem{Guy2023}
J.~{Guy}, \emph{et~al.}, {The Spectroscopic Data Processing Pipeline for the Dark Energy Spectroscopic Instrument}. \emph{\aj} \textbf{165}~(4), 144 (2023), \doi{10.3847/1538-3881/acb212}.

\bibitem{Carnall17}
A.~C. {Carnall}, {SpectRes: A Fast Spectral Resampling Tool in Python}. \emph{arXiv e-prints} arXiv:1705.05165 (2017), \doi{10.48550/arXiv.1705.05165}.

\bibitem{Zhu15}
G.~B. {Zhu}, \emph{et~al.}, {Near-ultraviolet Spectroscopy of Star-forming Galaxies from eBOSS: Signatures of Ubiquitous Galactic-scale Outflows}. \emph{\apj} \textbf{815}~(1), 48 (2015), \doi{10.1088/0004-637X/815/1/48}.

\bibitem{Martin13}
C.~L. {Martin}, \emph{et~al.}, {Scattered Emission from z \raisebox{-0.5ex}\textasciitilde 1 Galactic Outflows}. \emph{\apj} \textbf{770}~(1), 41 (2013), \doi{10.1088/0004-637X/770/1/41}.

\bibitem{Scarlata15}
C.~{Scarlata}, N.~{Panagia}, {A Semi-analytical Line Transfer Model to Interpret the Spectra of Galaxy Outflows}. \emph{\apj} \textbf{801}~(1), 43 (2015), \doi{10.1088/0004-637X/801/1/43}.

\bibitem{Carr18}
C.~{Carr}, C.~{Scarlata}, N.~{Panagia}, A.~{Henry}, {A Semi-analytical Line Transfer (SALT) Model. II: The Effects of a Bi-conical Geometry}. \emph{\apj} \textbf{860}~(2), 143 (2018), \doi{10.3847/1538-4357/aac48e}.

\bibitem{Lan18}
T.-W. {Lan}, H.~{Mo}, {The Circumgalactic Medium of eBOSS Emission Line Galaxies: Signatures of Galactic Outflows in Gas Distribution and Kinematics}. \emph{\apj} \textbf{866}~(1), 36 (2018), \doi{10.3847/1538-4357/aadc08}.

\bibitem{Rubin12}
K.~H.~R. {Rubin}, J.~X. {Prochaska}, D.~C. {Koo}, A.~C. {Phillips}, {The Direct Detection of Cool, Metal-enriched Gas Accretion onto Galaxies at z \raisebox{-0.5ex}\textasciitilde 0.5}. \emph{\apjl} \textbf{747}~(2), L26 (2012), \doi{10.1088/2041-8205/747/2/L26}.

\bibitem{Weldon23}
A.~{Weldon}, \emph{et~al.}, {The MOSDEF-LRIS survey: detection of inflowing gas towards three star-forming galaxies at z 2}. \emph{\mnras} \textbf{523}~(4), 5624--5634 (2023), \doi{10.1093/mnras/stad1615}.

\bibitem{Coleman24}
E.~{Coleman}, \emph{et~al.}, {Detection of Gas Inflow during the Onset of a Starburst in a Low-mass Galaxy at z = 2.45}. \emph{\apjl} \textbf{977}~(1), L23 (2024), \doi{10.3847/2041-8213/ad93d0}.

\bibitem{Burchett21}
J.~N. {Burchett}, \emph{et~al.}, {Circumgalactic Mg II Emission from an Isotropic Starburst Galaxy Outflow Mapped by KCWI}. \emph{\apj} \textbf{909}~(2), 151 (2021), \doi{10.3847/1538-4357/abd4e0}.

\bibitem{Zabl21}
J.~{Zabl}, \emph{et~al.}, {MusE GAs FLOw and Wind (MEGAFLOW) VIII. Discovery of a MgII emission halo probed by a quasar sightline}. \emph{\mnras} \textbf{507}~(3), 4294--4315 (2021), \doi{10.1093/mnras/stab2165}.

\bibitem{Moster2010}
B.~P. {Moster}, \emph{et~al.}, {Constraints on the Relationship between Stellar Mass and Halo Mass at Low and High Redshift}. \emph{\apj} \textbf{710}~(2), 903--923 (2010), \doi{10.1088/0004-637X/710/2/903}.

\bibitem{Maccio07}
A.~V. {Macci{\`o}}, \emph{et~al.}, {Concentration, spin and shape of dark matter haloes: scatter and the dependence on mass and environment}. \emph{\mnras} \textbf{378}~(1), 55--71 (2007), \doi{10.1111/j.1365-2966.2007.11720.x}.

\bibitem{Guseva2013}
N.~G. {Guseva}, Y.~I. {Izotov}, K.~J. {Fricke}, C.~{Henkel}, {The Mg II {\ensuremath{\lambda}}2797, {\ensuremath{\lambda}}2803 emission in low-metallicity star-forming galaxies from the SDSS}. \emph{\aap} \textbf{555}, A90 (2013), \doi{10.1051/0004-6361/201221010}.

\bibitem{Feltre18}
A.~{Feltre}, \emph{et~al.}, {The MUSE Hubble Ultra Deep Field Survey. XII. Mg II emission and absorption in star-forming galaxies}. \emph{\aap} \textbf{617}, A62 (2018), \doi{10.1051/0004-6361/201833281}.

\bibitem{Xu22a}
X.~{Xu}, \emph{et~al.}, {Tracing Ly{\ensuremath{\alpha}} and LyC Escape in Galaxies with Mg II Emission}. \emph{\apj} \textbf{933}~(2), 202 (2022), \doi{10.3847/1538-4357/ac7225}.

\bibitem{Prochaska11}
J.~X. {Prochaska}, D.~{Kasen}, K.~{Rubin}, {Simple Models of Metal-line Absorption and Emission from Cool Gas Outflows}. \emph{\apj} \textbf{734}~(1), 24 (2011), \doi{10.1088/0004-637X/734/1/24}.

\bibitem{pyqsofit}
H.~{Guo}, Y.~{Shen}, S.~{Wang}, {PyQSOFit: Python code to fit the spectrum of quasars}, Astrophysics Source Code Library, record ascl:1809.008 (2018).

\bibitem{Pandya21}
V.~{Pandya}, \emph{et~al.}, {Characterizing mass, momentum, energy, and metal outflow rates of multiphase galactic winds in the FIRE-2 cosmological simulations}. \emph{\mnras} \textbf{508}~(2), 2979--3008 (2021), \doi{10.1093/mnras/stab2714}.

\bibitem{Berg2025}
T.~A.~M. {Berg}, \emph{et~al.}, {Mapping the spatial extent of H I-rich absorbers using Mg II absorption along gravitational arcs}. \emph{\aap} \textbf{693}, A200 (2025), \doi{10.1051/0004-6361/202452199}.

\bibitem{Baldwin1981}
J.~A. {Baldwin}, M.~M. {Phillips}, R.~{Terlevich}, {Classification parameters for the emission-line spectra of extragalactic objects.} \emph{\pasp} \textbf{93}, 5--19 (1981), \doi{10.1086/130766}.

\bibitem{Pucha24}
R.~{Pucha}, \emph{et~al.}, {Tripling the Census of Dwarf AGN Candidates Using DESI Early Data}. \emph{arXiv e-prints} arXiv:2411.00091 (2024), \doi{10.48550/arXiv.2411.00091}.

\bibitem{Juneau11}
S.~{Juneau}, M.~{Dickinson}, D.~M. {Alexander}, S.~{Salim}, {A New Diagnostic of Active Galactic Nuclei: Revealing Highly Absorbed Systems at Redshift >0.3}. \emph{\apj} \textbf{736}~(2), 104 (2011), \doi{10.1088/0004-637X/736/2/104}.

\bibitem{Juneau14}
S.~{Juneau}, \emph{et~al.}, {Active Galactic Nuclei Emission Line Diagnostics and the Mass-Metallicity Relation up to Redshift z \raisebox{-0.5ex}\textasciitilde 2: The Impact of Selection Effects and Evolution}. \emph{\apj} \textbf{788}~(1), 88 (2014), \doi{10.1088/0004-637X/788/1/88}.

\end{thebibliography}

%
%
%
%
%
%


\section*{Acknowledgments}
The authors thank the anonymous reviewers for their insightful and constructive comments, which have substantially contributed to improving the quality and clarity of the manuscript.
This research used data obtained with the Dark Energy Spectroscopic Instrument (DESI). DESI construction and operations is managed by the Lawrence Berkeley National Laboratory. This material is based upon work supported by the U.S. Department of Energy, Office of Science, Office of High-Energy Physics, under Contract No. DE–AC02–05CH11231, and by the National Energy Research Scientific Computing Center, a DOE Office of Science User Facility under the same contract. Additional support for DESI was provided by the U.S. National Science Foundation (NSF), Division of Astronomical Sciences under Contract No. AST-0950945 to the NSF’s National Optical-Infrared Astronomy Research Laboratory; the Science and Technology Facilities Council of the United Kingdom; the Gordon and Betty Moore Foundation; the Heising-Simons Foundation; the French Alternative Energies and Atomic Energy Commission (CEA); the National Council of Humanities, Science and Technology of Mexico (CONAHCYT); the Ministry of Science and Innovation of Spain (MICINN), and by the DESI Member Institutions: www.desi.lbl.gov/collaborating-institutions. The DESI collaboration is honored to be permitted to conduct scientific research on I’oligam Du’ag (Kitt Peak), a mountain with particular significance to the Tohono O’odham Nation. Any opinions, findings, and conclusions or recommendations expressed in this material are those of the author(s) and do not necessarily reflect the views of the U.S. National Science Foundation, the U.S. Department of Energy, or any of the listed funding agencies.

\paragraph*{Funding:}

This work was supported by the National Key R\&D Program of China (grant Nos. 2022YFA1602902 and 2023YFA1607804), the National Natural Science Foundation of China (NSFC; grant Nos. 12473008, 12192224, 12120101003 and 12373010) and CAS Project for Young Scientists in Basic Research, Grant No. YSBR-062.

\paragraph*{Author contributions:}

Conceptualization: EW, HW
Methodology: HY, ZC, EW
Investigation: HY, ZH
Visualization: HY, EW, HW 
Supervision: EW
Writing–original draft: HY
Writing–review \& editing: HY, EW, CP, HZ, HW, CL, CJ, CM, XK
Validation: HY, EW
Formal analysis: HY, EW, ZH
Funding acquisition: EW, HZ, HW
Data curation: HY
Software: HY
Project administration: EW
Resources: EW

\paragraph*{Competing interests:}

There are no competing interests to declare.

\paragraph*{Data, code and materials availability:}

All data and code needed to evaluate and reproduce the results in the paper are present in the paper and/or the Supplementary Materials.
This study did not generate new materials.
\subsection*{Supplementary materials}
Supplementary Text\\
Figs. S1 to S8\\
References \textit{(75-\arabic{enumiv})}\\ 


\newpage


\renewcommand{\thefigure}{S\arabic{figure}}
\renewcommand{\thetable}{S\arabic{table}}
\renewcommand{\theequation}{S\arabic{equation}}
\renewcommand{\thepage}{S\arabic{page}}
\setcounter{figure}{0}
\setcounter{table}{0}
\setcounter{equation}{0}
\setcounter{page}{1} 


\begin{center}
\section*{Supplementary Materials for\\ \scititle}


Haoran~Yu,
Enci~Wang$^{\ast}$,
Zeyu~Chen,
C\'eline~P\'eroux,\\
Hu~Zou,
Zhicheng~He,
Huiyuan~Wang,
Cheqiu~Lyu,\\
Cheng~Jia,
Chengyu~Ma,
Xu~Kong\\
\small$^\ast$Corresponding author. Email: ecwang16@ustc.edu.cn

\end{center}

\subsubsection*{This PDF file includes:}
Supplementary Text\\
Figures S1 to S8


\newpage


\section*{Supplementary Text}

\subsection*{An Alternative Approach: Modelling  Mg\,\textsc{ii} Emission}

In the main text, we directly use the \textsc{Boxcar} method to extract the outflow properties, namely $\mathrm{EW}_{\rm out}$ and $v_{\rm out}$, as this approach is nearly model independent and highly stable. This approach implicitly assumes that the Mg\,\textsc{ii} emission predominantly arises from the optically thick ISM gas, as observed in some low-mass galaxies \cite{Guseva2013, Feltre18, Xu22a}. 
However, this assumption may not always hold. We therefore present, in this section, an alternative method to model the Mg~\textsc{ii} emission and to extract the outflow signatures by subtracting this emission component.

\begin{figure}
    \centering
    \includegraphics[width=0.8\textwidth]{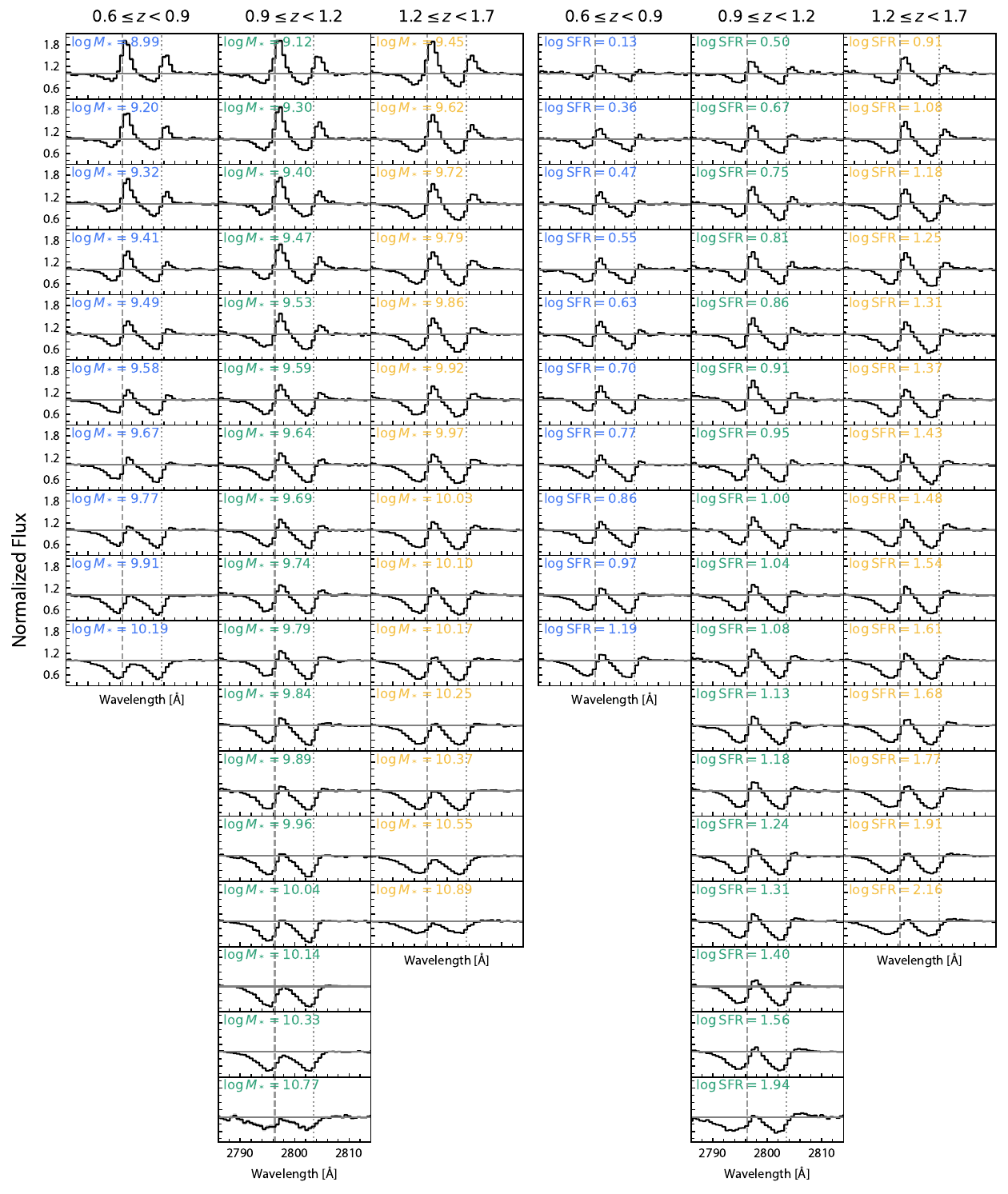}
    \caption{
    \textbf{The stacked galaxy spectra in bins of $M_*$ and $\rm SFR$.}
    Each spectrum is derived by stacking 10,000 spectra, corresponding to the data points in Figure~\ref{fig:evolution} in the main part.
    The left group of panels show the spectra in bins of  $M_*$, with increasing $M_*$ from top to bottom, while the right group of panels show the spectra in bins of $\rm SFR$.
    Blue, green and yellow represents samples with $0.6\leq z<0.9$, $0.9\leq z<1.2$ and $1.2\leq z<1.7$, respectively.
    The uncertainties of the spectra are small thus are not presented.
    }
    \label{fig:outflow_spec}
\end{figure}

As shown in Figure \ref{fig:outflow_spec}, the stacked profiles of galaxies with $M_*\lesssim10^{10}~\rm M_\odot$ present some P-Cygni-like features with slightly red-shifted emission lines, which may come from the scattered light from an expanding shell~\cite{Prochaska11, Martin13, Scarlata15, Carr18}.
We therefore present an alternative method to account for the visible emission by decomposing the stacked Mg~\textsc{ii} profiles into multiple absorption and emission components. Given the strong model dependence and sensitivity to assumptions inherent in emission-line decomposition, we place this alternative analysis in the supplementary text.

The intensity of an absorption line can be described as 
\begin{equation}
    I_{\rm abs}(\lambda)=1-C+Ce^{-\tau(\lambda)},
\end{equation}
where $C$ is the covering factor and $\tau(\lambda)$ is optical depth varying with wavelength. 
For simplicity, $\tau(\lambda)$ is parameterized as a Gaussian:
\begin{equation}
    \tau(\lambda) = \tau_0 e^{-(\lambda - \lambda_0)^2 / (\lambda_0b/c)^2}.
\end{equation}
In this expression, $\tau_0$ and $\lambda_0$ are the central optical depth and the central wavelength, $b$ is the Doppler parameter and $c$ is the speed of light. 
Thus an absorption profile with known line center ($\lambda_0$) can be described with 3 parameters: $C$, $\tau_0$ and $b$. 
The intensity of an emission line can be formulated as a Gaussian:
\begin{equation}
    I_{\rm em}(\lambda)=I_0e^{-(\lambda-\lambda_0)^2/(\lambda_0\sigma/c)^2},
\end{equation}
where $I_0$ is the peak of the profile, $\lambda_0$ is the central wavelength and $\sigma$ is the width of the Gaussian.
An emission profile with known center can be well parameterized with 2 parameters: $I_0$ and $\sigma$.

The profile of Mg\,\textsc{ii} doublet is decomposed into 6 components, including 2 systemic absorption profiles, 2 blueshifted absorption profiles and 2 emission profiles representing the scattered emission. 
While the systemic absorption profiles have their velocities fixed at $v=0 \ {\rm km \ s^{-1}}$, the outflow components and the emission components have variable central velocities. 
Note that this assumption is different from the main analysis using \textsc{Boxcar} method, where we assume that the emission profiles are symmetrical with respect to $v=0 \ {\rm km \ s^{-1}}$.
Therefore, 
20 parameters in total are needed to describe the Mg\,\textsc{ii} profile.
To alleviate the degeneracy in the fitting process, we bind some parameters that are expected to be physically connected.
\begin{itemize}
    \item For the systemic absorption and the blueshifted absorption, we assume the profiles are identical between Mg\,\textsc{ii}\,$\lambda$2796 and Mg\,\textsc{ii}\,$\lambda$2803.
    \item Based on the emission line ratios of approximately 2:1, the scattering shell is presumably optically thin. Thus we assume $I_0\equiv I_0({\rm Mg\,\textsc{ii}\,\lambda 2796}) = 2I_0(\rm Mg\,\textsc{ii}\,\lambda 2803)$.
    \item We assume the emission line doublets have the same width, $\sigma$.
    \item We assume the central velocities of the two emission lines are identical ($v_{\rm em}$), and the central velocities of the two blueshifted absorption lines are identical ($v_{\rm blue}$).
\end{itemize}
Thus we use 10 parameters ($C_{\rm sys}$, $\tau_{0, \rm sys}$, $b_{\rm sys}$, $C_{\rm blue}$, $\tau_{0, \rm blue}$, $b_{\rm blue}$, $I_0$, $\sigma$, $v_{\rm em}$ and $v_{\rm blue}$) to describe the Mg\,\textsc{ii} profile, consisting of 6 components ($I_{\rm sys, 2796}$, $I_{\rm blue, 2796}$, $I_{\rm em, 2796}$, $I_{\rm sys, 2803}$, $I_{\rm blue, 2803}$ and $I_{\rm em, 2803}$). 
Considering the clumpiness of the cool gas traced by Mg\,\textsc{ii}, the scattered emission is not likely to be absorbed, therefore the spectrum is parameterized as 
\begin{equation}
    I = I_{\rm sys, 2796}\cdot I_{\rm blue, 2796} \cdot I_{\rm blue, 2803}\cdot I_{\rm sys, 2803}+I_{\rm em, 2796} +I_{\rm em, 2803}
\end{equation}

\begin{figure}
    \centering
    \includegraphics[width=0.7\linewidth]{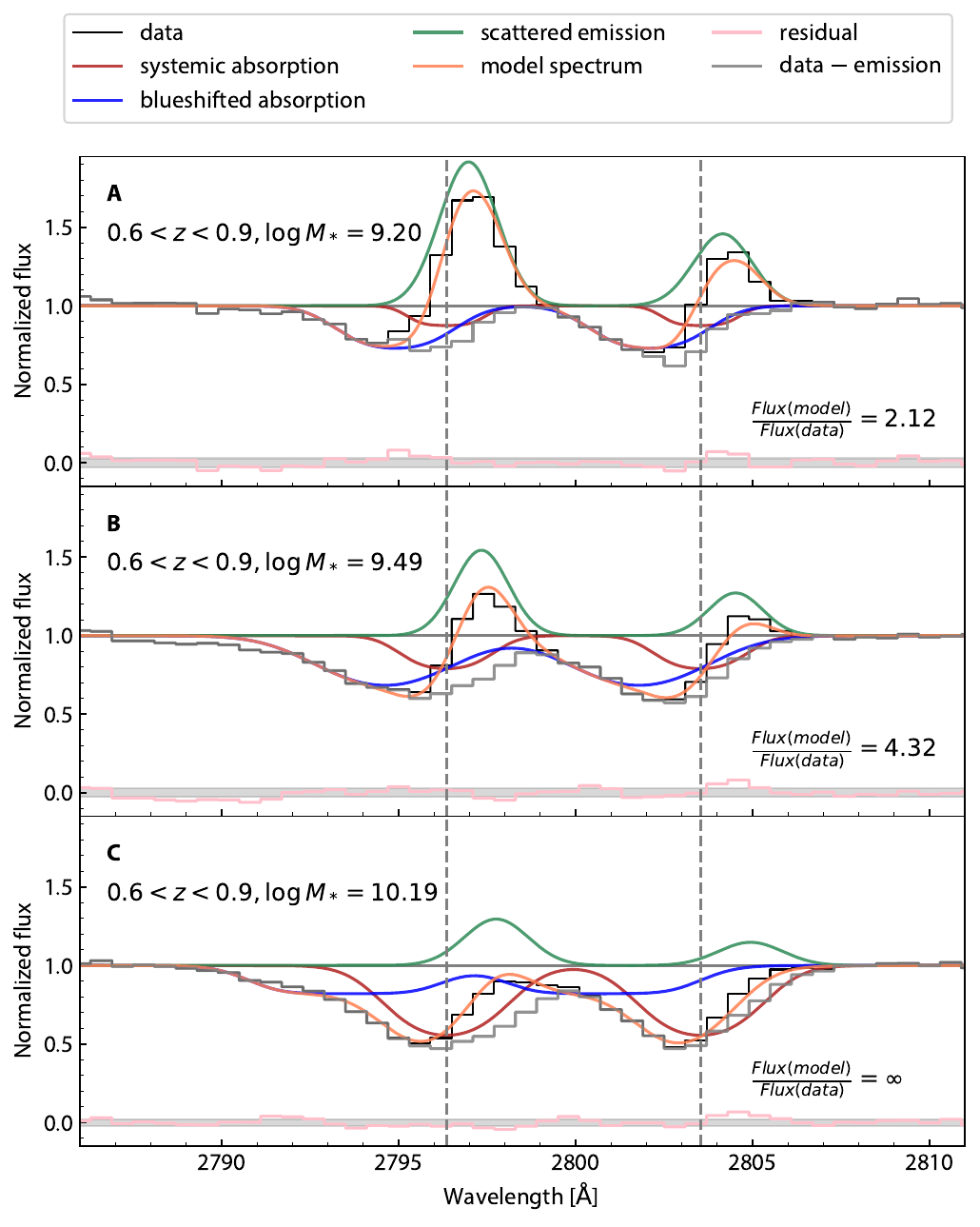}
    \caption{
    \textbf{Example multicomponent fitting results.}
    \textbf{(A)}
    Composite spectrum of low mass galaxies in which the scattered emission is significant.
    The black curve represents data from the composite spectra.
    The red, blue and green curve represent systemic absorption, blueshifted absorption and emission components, respectively.
    The orange curve represents the model spectrum, and the pink curve represents residual which is data $-$ model spectrum.
    The shaded region around flux=0 indicates 1$\sigma$ uncertainty inferred from MCMC fitting.
    \textbf{(B)}
    Composite spectrum of medium mass galaxies in which the emission is moderate.
    \textbf{(C)}
    Composite spectrum of massive galaxies, displaying no emission.
    }
    \label{fig:modelfit}
\end{figure}

For a composite spectrum we run Monte Carlo Markov Chain fitting with flat prior of $0<\tau_{0, \rm sys},\tau_{0, \rm out}<10$, $0<b_{\rm sys}<300 \ {\rm km \ s^{-1}}$, $0<b_{\rm blue}<500 \ {\rm km \ s^{-1}}$, $0< C_{\rm sys}, C_{\rm blue}\leq 1$, $0\leq I_0<2$, $0<\sigma<300 \ {\rm km \ s^{-1}}$, $0<v_{\rm blue}<500 \ {\rm km \ s^{-1}}$, $-200<v_{\rm em}<200 \ {\rm km \ s^{-1}}$.
For galaxies with $M_* \gtrsim 10^{10}\,\rm M_\odot $, emission features are largely absent in the stacked spectra, suggesting that the fitted emission components in this regime may not be physically meaningful. Nevertheless, we perform the decomposition to all composite spectra across the full range of stellar mass and SFR bins for consistency.
Figure~\ref{fig:modelfit} shows several examples of model fitting in different conditions, from prominent to null visible emission.
Although the model is able to reproduce the data in nearly all cases, we caution that when fitting spectra showing weak or null emission, the emission components are poorly constrained and can be overfit.
We calculate the ratio between the integrated emission flux in the model and the sum of positive emission flux in the data ($\rm\frac{Flux(model)}{Flux(data)}$) to quantify the reliability of the model.
A larger ratio indicates more severe degeneracy between emission and absorption, implying that the intensity of the emission lines could be overpredicted.

The assumption that the absorption profiles of the doublets are identical is a compromise with the insufficient spectroscopic resolution.
We also test with the assumption that $\tau(\rm Mg\,\textsc{ii}\,\lambda2796)=2\tau(\rm Mg\,\textsc{ii}\,\lambda2803)$, which is adopted by many works to fit the doublets at the same time, but it returns systematically higher standard error in residual ($\gtrsim10\%$) and is not adopted.
It could be because of the optical thickness of the Mg\,\textsc{ii} cloud, making Gaussian inappropriate to describe $\tau(\lambda)$.
We note that the primary aim of this fitting analysis is to subtract the emission component using an approach different from the one used in the main text.
Therefore, the fitted systemic and blueshifted absorption only represents the shape of the profile and does not physically indicate ISM and outflow components.

\begin{figure}
    \centering
    \includegraphics[width=0.6\linewidth]{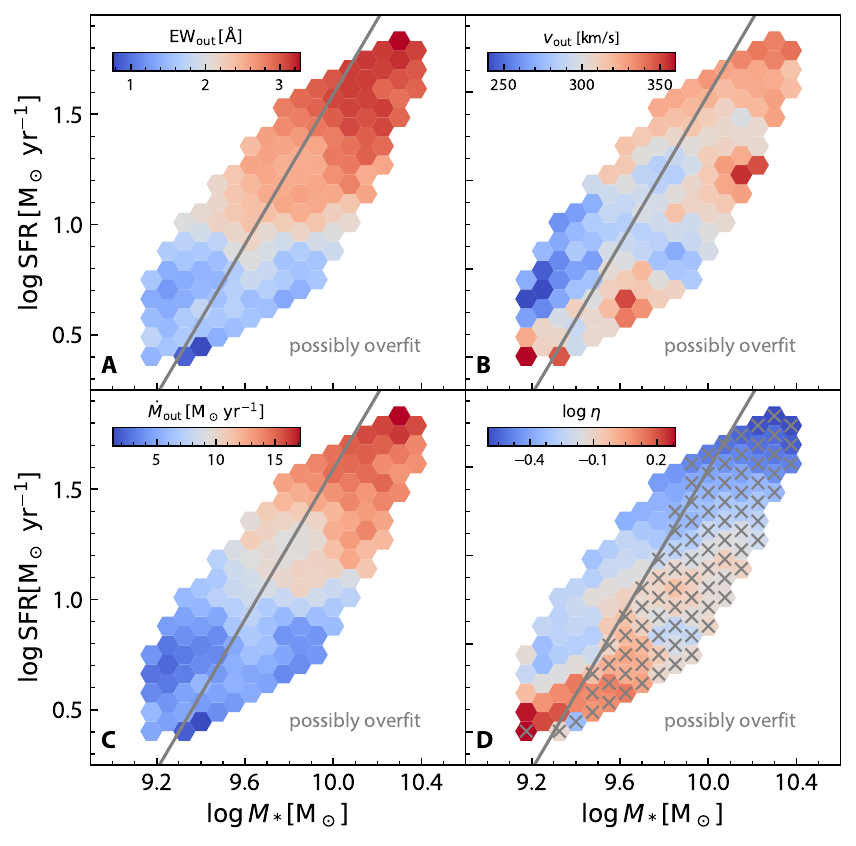}
    \caption{
    \textbf{Outflow properties of samples with $0.6<z<1.7$ on the $M_*$-$\rm SFR$ diagram, after removing the fitted emission components.}
    \textbf{(A)}
    $\rm EW_{out}$ map.
    The binning is identical to that in Figure~\ref{fig:map} in the maintext, but the outflow properties are derived by first removing the fitted emission lines and then applying the \textsc{Boxcar} method.
    The bins at the right side of the grey line are derived with emission line flux ratios of $\rm \frac{Flux(model)}{Flux(data)}>3$, indicating that the emission component could be overfit and the results are unreliable.
    \textbf{(B)}
    $v_{\rm out}$ map.
    \textbf{(C)}
    $\dot{M}_{\rm out}$ map.
    \textbf{(D)}
    $\eta$ map, where the grey diagnal crosses denotes the exact bins where the emission is potentially overfit.
    }
    \label{fig:map_noem}
\end{figure}

After accounting for the emission with off-center Gaussians with flux ratio of 2:1, we apply \textsc{Boxcar} method to pure absorption profiles, which is data $-$ emission as indicated in Figure~\ref{fig:modelfit}, to derive outflow properties.
We first examine whether removing the fitted emission affect the mapping of outflow properties on SFMS, as shown in Figure~\ref{fig:map_noem}.
Compared with Figure~\ref{fig:map}, we find that the removement of emission causes decrease in $\rm EW_{out}$ and increase in $v_{\rm out}$.
The color gradients in the hexogonal bins are slightly different from the ones presented in the maintext, but the overall trends are essentially consistent: $\rm EW_{out}$ is larger in bins with higher SFR, while $v_{\rm out}$ is higher at massive galaxies.
We note that the correlation between $v_{\rm out}$ and $M_*$ is not as stable as in the main analysis, and the displayed overall fluctuation in $v_{\rm out}$ map is strong. 
This is attributed to the extra uncertainty introduced in the model fitting process.
The derived $\dot{M}_{\rm out}$ and $\eta$ are marginally lower than in Figure~\ref{fig:map},
although displaying higher fluctuation.
We denote the bins with $\rm\frac{Flux(model)}{Flux(data)}>3$ as unreliable results, as the emission lines are possibly overfit.
These bins are located essentially below the grey line as displayed in Figure~\ref{fig:map_noem}, with high $M_*$ and low SFR.

\begin{figure}
    \centering
    \includegraphics[width=\linewidth]{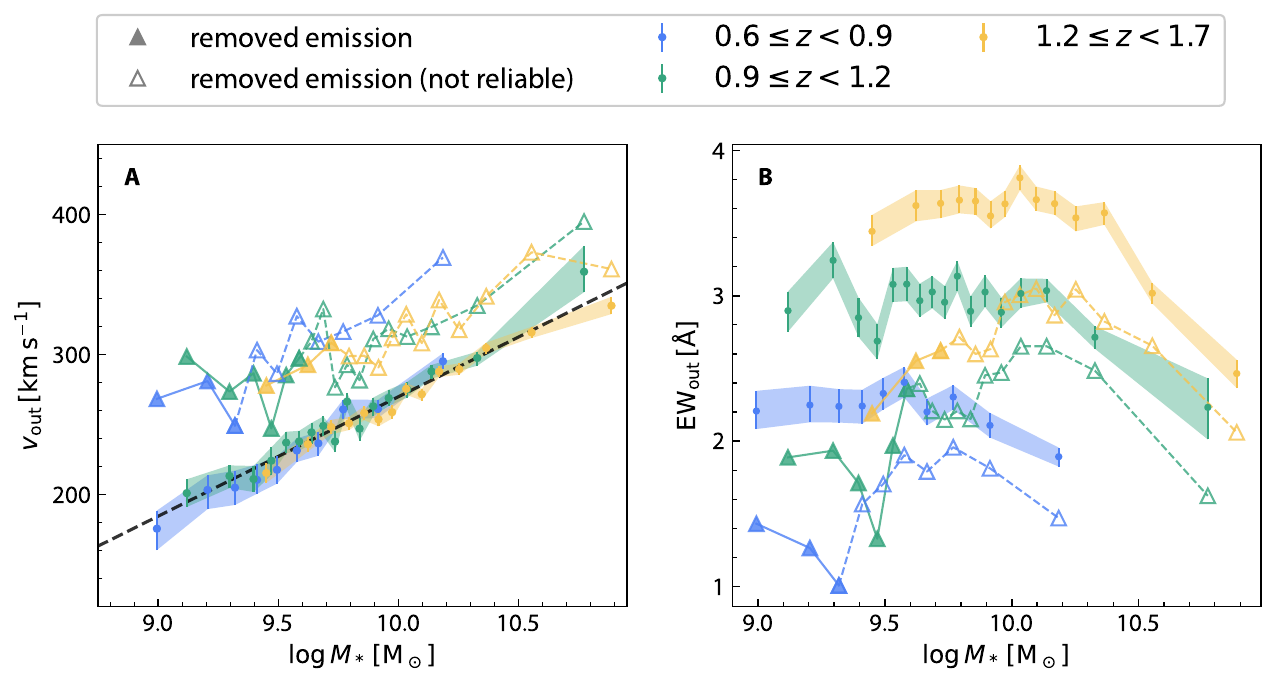}
    \caption{
    \textbf{Evaluating the effect of removing fitted emission on $v_{\rm out}$ and $\rm EW_{out}$.}
    \textbf{(A)}
    $v_{\rm out}$ vs. $\log M_*$ in 3 redshift bins.
    Blue, green and yellow color represent samples with $0.6<z<0.9$, $0.9<z<1.2$ and $1.2<z<1.7$.
    The dots and the shaded region represent the result and the corresponding 1$\sigma$ uncertainty from Figure~\ref{fig:evolution} in the maintext.
    The black dashed line represents the empirical correlation described in Equation~\ref{eq:vout}.
    The triangles represent results derived by first removing fitted emission components, with the open ones denoting those with emission line flux ratios of $\rm \frac{Flux(model)}{Flux(data)}>3$, which potentially suffers from overpredicted emission.
    \textbf{(B)}
    $\rm EW_{out}$ vs. $\log M_*$ in 3 redshift bins.
    }
    \label{fig:scaling}
\end{figure}

In Figure~\ref{fig:scaling} we evaluate the effect of removing fitted emission on $v_{\rm out}$ and $\rm EW_{out}$.
After removing the fitted redshifted emission, we find systematic increase in $v_{\rm out}$ in all bins.
However, the bins with relatively reasonable emission subtraction (with $\rm\frac{Flux(model)}{Flux(data)}<3$) is limited within low-mass galaxies, where the emission is prominent.
This strengthens that outflows in low-mass star-forming galaxies can effectively escape from the host DMH.
On the other hand, the outflow velocities of massive galaxies with $M_*\gtrsim10^{9.5}~\rm M_\odot$ are potentially overestimated and directly applying \textsc{Boxcar}  is more appropriate. 

This model fitting method is also subject to multiple sources of uncertainties.
The most substantial concern is the degeneracy between emission and absorption profiles in the spectral region bracketed by the doublet lines.
While spectra exhibiting weak visible emission have been flagged as unreliable fits, the robustness of the fitting in other bins is also to be questioned.
Furthermore, the assumption that the blueshifted absorption and the emission can be represented by a Gaussian is also idealized, given that the line profile could possibly be red-skewed or otherwise asymmetric. 
Although we consider this approach a plausible means of interpreting the complicated Mg\,\textsc{ii} spectroscopic profile, its reliance on specific model assumptions and the additional uncertainties it introduces lead us to treat it as complementary to our primary analysis.

\subsection*{Mg\,\textsc{ii} Absorption in the CGM}

As demonstrated in the main text, we find that outflow velocities decline rapidly as they propagate in massive galaxies ($M_*>10^{10}\,{\rm M_{\odot}}$). 
This behavior may leave a distinct imprint on the metal distribution within the CGM, potentially differentiating low- and high-mass galaxies.
To explore this further, we analyze the total equivalent width of Mg\,\textsc{ii} doublets, $\rm EW$(Mg\,\textsc{ii}), in the CGM using stacked background quasar spectra. 
The methodology for this CGM analysis differs somewhat from the approach adopted in our primary study, as summarized below (see Chen et al.~\cite{Chen24} for a comprehensive description).

Using the DESI EDR catalog, we identify quasar-ELG pairs by cross-matching their coordinates and selecting those with a projected impact parameter of $<150\,\rm kpc$.
To mitigate contamination from quasar-associated absorption features, we require that the quasar redshifts exceed those of the foreground galaxies by $\Delta z>0.02$.
To investigate the azimuthal dependence of Mg\,\textsc{ii} absorption, we further require that the galaxies have position angle measurements from DESI legacy imaging surveys. 
We first fit the intrinsic quasar spectra using \texttt{PyQSOFit}~\cite{pyqsofit} for each individual spectrum.
The observed spectra are then normalized by dividing them by their best-fit quasar model spectrum.
To further suppress high-frequency noise, we apply a median filter to the normalized spectra, excluding the wavelength regions containing Mg\,\textsc{ii} absorption to avoid biasing the smoothing process. 
Finally, we stack the processed spectra in the rest-frame of the foreground galaxies, generating a composite spectrum by taking the median flux at each wavelength.

\begin{figure}
\centering
    \includegraphics[width=0.8\textwidth]{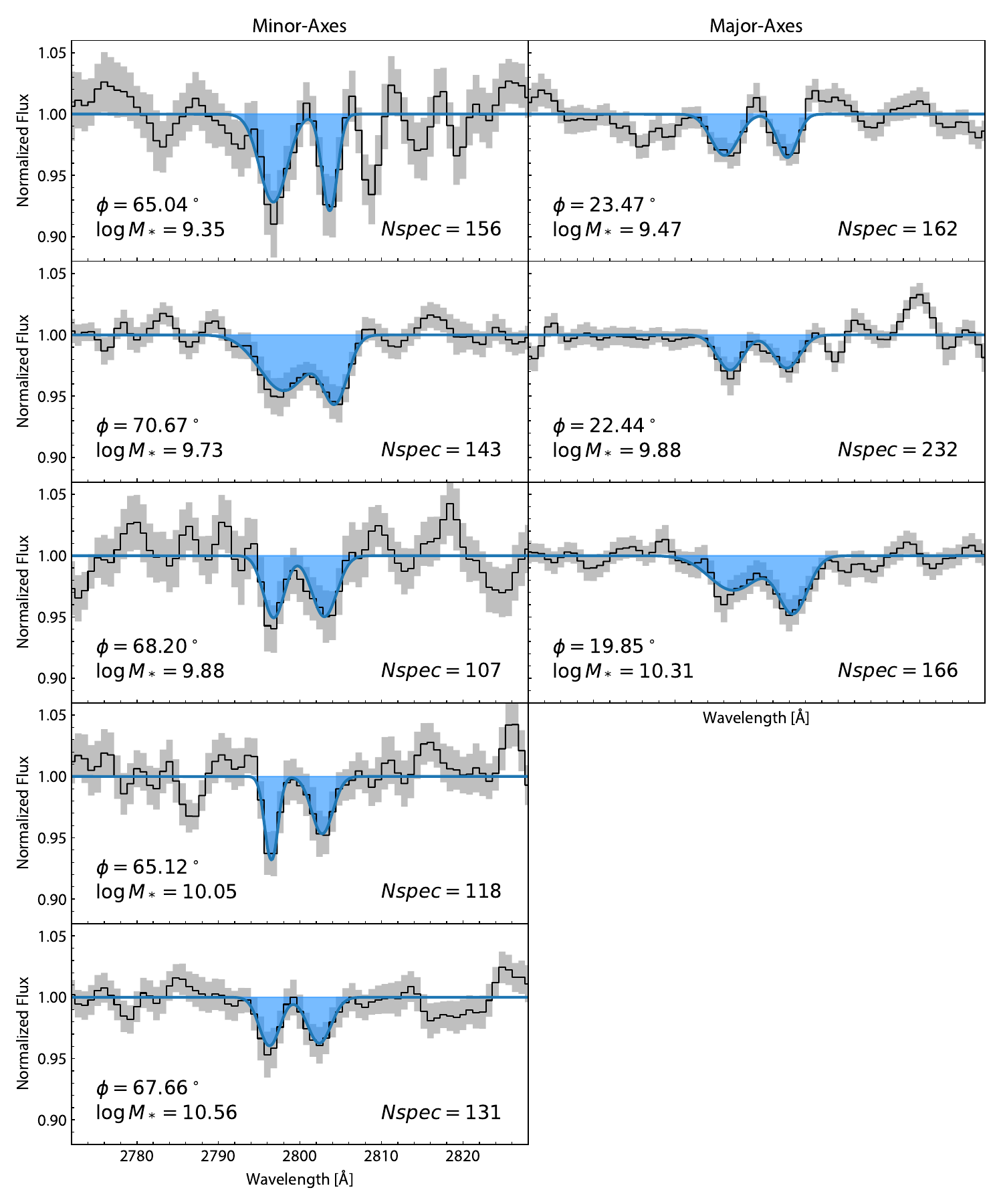}
    \caption{
    \textbf{The stacked spectra of 20-150\,kpc CGM around galaxies.}
    The left column marked ``Minor-Axes'' indicates the CGM is within $45^\circ$ of the minor-axes measured from the galactic center.
    The right column marked ``Major-Axes'' indicates the CGM is within $45^\circ$ of the major-axes.
    The black curve represents the median spectra and the grey shaded area indicates error derived from bootstrapping.
    The blue curve represents the fitted double Gaussian which is only for the purpose of illustration, while the EW of the absorption is calculated through integrating across the wavelengths within 3$\sigma$ of the fitted Gaussian model.
    In each panel, the number of spectra, the mean position angle relative to the major axis ($\phi$), and the mean stellar mass within each bin are indicated.
    Each composite spectrum is smoothed with a gaussian kernel with sigma of $1\,\mathring{\rm A}$.
    }
    \label{fig:cgm_spec}
\end{figure}

The stacked Mg\,\textsc{ii} absorption spectra of the CGM are shown in Figure~\ref{fig:cgm_spec}, which corresponds to the data points in the bottom panel of Figure~\ref{fig:gravity} in the main text.
To quantify the absorption strength, we fit the Mg\,\textsc{ii} doublet using a double-Gaussian profile.
The Mg\,\textsc{ii} total equivalent width, $\rm EW(Mg\,\textsc{ii})$, is derived by integrating the absorption flux over the 3$\sigma$ spectral windows centered on the best-fit Gaussian profiles \cite{Chen24}.
Uncertainties in the EW measurements are derived from 100 bootstrap iterations.
Some unexpected features appear beyond the Mg\,\textsc{ii} doublets, which could originate from the smoothing kernel applied to the spectra, given the limited stacking sample size and thus low S/N of the composite spectra.

\begin{figure}
    \centering
    \includegraphics[width=0.6\linewidth]{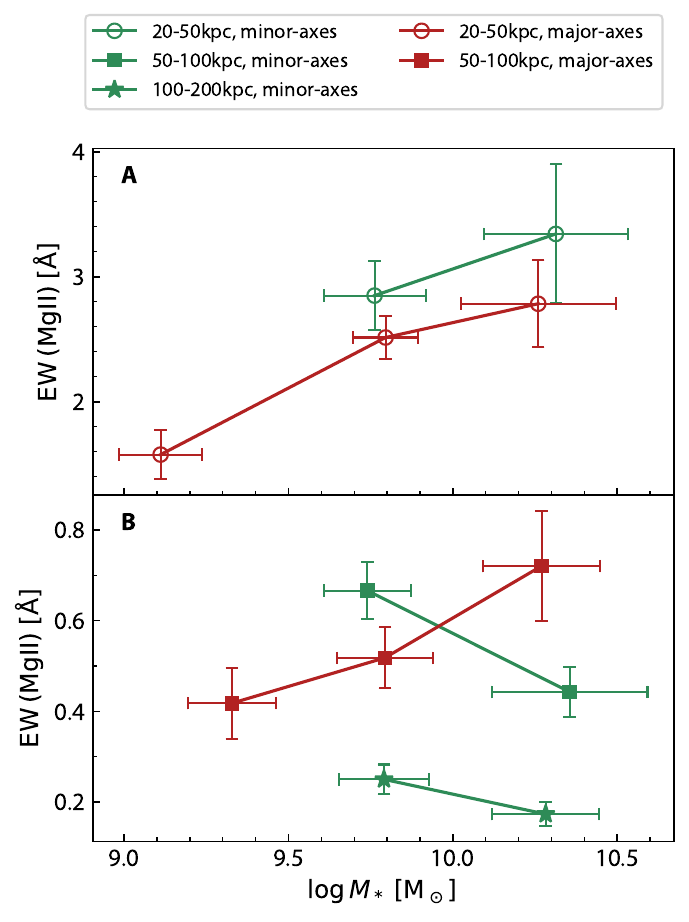}
    \caption{
    \textbf{EW(Mg\,\textsc{ii}) vs. $\log M_*$ in impact parameter bins.}
    \textbf{(A)}
    The circles represent EW(Mg\,\textsc{ii}) in the 20-50~kpc CGM. 
    The green (red) symbols represent CGM at the minor (major)-axes of the galaxy.
    \textbf{(B)}
    The squares and the stars represent EW(Mg\,\textsc{ii}) in the 50-100~kpc CGM and 100-200~kpc CGM. 
    Note that the scales in $y$-axes are different between the two panels.
    }
    \label{fig:cgm_bins}
\end{figure}

Based on our finding that CGM Mg\,\textsc{ii} absorption is suppressed along the minor-axes of galaxies at $M_*\gtrsim10^{10}\,\rm M_\odot$, we proceed to investigate the azimuthal dependence of CGM absorption across different impact parameter bins.
As illustrated in Figure~\ref{fig:cgm_bins}, within the 20-50~kpc impact parameter range --- where outflows from massive galaxies are expected to reach as indicated in Figure~\ref{fig:gravity} --- EW(Mg\,\textsc{ii}) increases with $M_*$ in both minor and major-axes.
This trend can be primarily attributed to the fact that more massive halos generally host larger baryonic reservoirs.
Notably, the enhancement of EW(Mg\,\textsc{ii}) along the minor-axes with respect to the major-axes suggests that the outflows enrich the CGM preferentially within a biconical geometry.
At impact parameter beyond 50 kpc as shown in the bottom panel of Figure~\ref{fig:cgm_bins}, EW(Mg\,\textsc{ii}) is substantially lower than at 20-50 kpc.
In the minor-axes of both 50-100 kpc and 100-200 kpc regions, we observe a decline in EW(Mg\,\textsc{ii}) with increasing $M_*$, consistent with the predictions from our gravitational potential modelling.
These results further support our conclusion that stellar feedback driven outflows get recycled in massive galaxies as a result of the deeper gravitational potential.

\subsection*{The Mass-Loading Factor}

Here we discuss the factors that could cause inconsistency in $\eta$ between simulations and observations in panel B of Figure~\ref{fig:eta}.
First, the observational results are derived from heterogenous samples while using slightly different methodologies.
Schroetter et al. 2019~\cite{Schroetter19} probed outflows along the minor axes of $z\sim 1$ galaxies using Mg\,\textsc{ii} absorption line in the spectra of background quasars, and estimated $\eta$ by assuming bi-conical geometry of the outflowing material.
Perrotta et al. 2023~\cite{Perrota23} analyzed a sample of extreme starburst galaxies at $z=0.4$-$0.7$ using down-the-barrel Mg\,\textsc{ii} absorption, naturally yielding much higher $\eta$ than the result of this work. 
Reichardt Chu et al. 2025~\cite{ReichardtChu25} detected ionized outflows in 10 low redshift starburst galaxies, by decomposing H$\beta$ and [O\,\textsc{III}]\,$\lambda$5007 emission lines. 
Although their result appears consistent with our predictions, we caution that the cool outflows traced by Mg\,\textsc{ii} may not be physically correlated with the warm, ionized outflows.

Compared with Illustris~\cite{Nelson15}, IllustrisTNG~\cite{Nelson19} and FIRE~\cite{Angles17} at similar redshifts, our data shows similar slope but is lower by 0.5-1.0~dex.
$\eta$ from EAGLE~\cite{Mitchell20} is close to our estimates for $M_*\lesssim10^{10}~\rm M_\odot$ galaxies, but deviates towards higher values for massive systems, which could be an effect of the AGN feedback implemented in the simulation.
However, we note several caveats in calculating the outflow rate, including the emipirical conversion from $\rm EW_{\lambda 2796}$ to $N_{\rm HI}$, the simplified geometry of the outflows, and the assumed launching radius of cool gas.
These parameter assumptions introduce systematic error in estimating the mass loading factor, making comparison between different works difficult.
Note that the physical conditions adopted by different simulations varies to some extent.
While in IllustrisTNG the mass loading is calculated for particles with positive radial velocity at a distance of 10~kpc~\cite{Nelson19}, in FIRE it is estimated by fitting particle-tracking based measurements~\cite{Pandya21}, and EAGLE considers mass loading on ISM scale.
We note that a persistent tension remains between the state-of-the-art simulations and observations, highlighting the need for further efforts to reconcile the two from both sides.

In addition, the assumption of a unity covering fraction adopted throughout the main analysis may lead to underestimation of $\dot{M}_{\rm out}$ and $\eta$ to some extent.
The absorption detected against an extended background source may be only partially covering, attributable to the inhomogeneous, clumpy structure of the intervening absorbing medium~\cite{Berg2025}.
However, ${\rm EW}_{\lambda2796}$ in Equation~\ref{eq:nhi} is initially measured from quasar background spectra, for which the covering fraction is unity.
To account for a non-unity covering fraction ($C_{\rm f}$), the inferred $N_{\rm HI}$ should be rescaled by a factor of $C_{\rm f}^{-\alpha}$, where $\alpha=1.69\pm 0.13$. 
Then $\dot{M}_{\rm out}$ is multiplied by an additional factor of $C_{\rm f}$, rendering an overall dependence on $C_{\rm f}^{1-\alpha}$. 
Under the assumption of $C_{\rm f} = 0.5$, $\dot{M}_{\rm out}$, and $\eta$ would increase by 0.2~dex.
We again strengthen that the estimates of $\dot{M}_{\rm out}$ and $\eta$ are subject to substantial uncertainties and should therefore be considered as secondary outcomes of this work.

\subsection*{Active Galactic Nuclei Contamination}

This study does not \textit{a priori} exclude AGN contributions in spectral stacking, given the difficulty in identifying AGN from individual spectra.
Two-dimensional emission line diagnostic diagrams, such as the Baldwin–Phillips–Terlevich (BPT) diagram~\cite{Baldwin1981} employs [O\,\textsc{iii}\,$\lambda$5007]/H$\beta$ versus [N\,\textsc{ii}\,$\lambda$6583]/H$\alpha$ to distinguish the primary source of ionization in galaxies. 
Applying to DESI data (with spectral coverage of 3600-9800$\rm \mathring{A}$ in observational frame), the BPT diagram can only identify AGN at $z<0.45$~\cite{Pucha24}.
Even when adopting the ``Mass-Excitation'' (MEx) diagnostic diagram~\cite{Juneau11, Juneau14}, which relies on [O\,\textsc{iii}\,$\lambda$5007]/H$\beta$, the construction of an AGN-clean sample is restricted to $z\lesssim0.9$.
Furthermore, robust AGN diagnosis also demand sufficiently high S/N for the relevant emission line measurements in individual spectra, which would exclude many low-S/N sources and consequently decrease the statistical significance of our stacking analysis.

Given the aforementioned limitations, we explicitly evaluate potential influence of AGN in this section. 
We employ the MEx diagnostic diagram to identify AGNs, focusing on the subset of galaxies with $z\lesssim0.9$ where H$\beta$ and [O\,\textsc{iii}]\,$\lambda$5007 emission lines are accessible.
By applying SNR thresholds of $\rm SNR(H\beta)>5$ and $\rm SNR([O\,\textsc{iii}]\,\lambda 5007)>10$, we retain 40,079 galaxies, with 7,114 (17.7\% of this subset) classified as AGNs.
As the decline in AGN prevalence with redshift for low-mass galaxies has been observed~\cite{Pucha24}, the AGN fraction in higher-redshift ($z>1$) galaxies is likely no larger than this value.

\begin{figure}
    \centering
    \includegraphics[width=\textwidth]{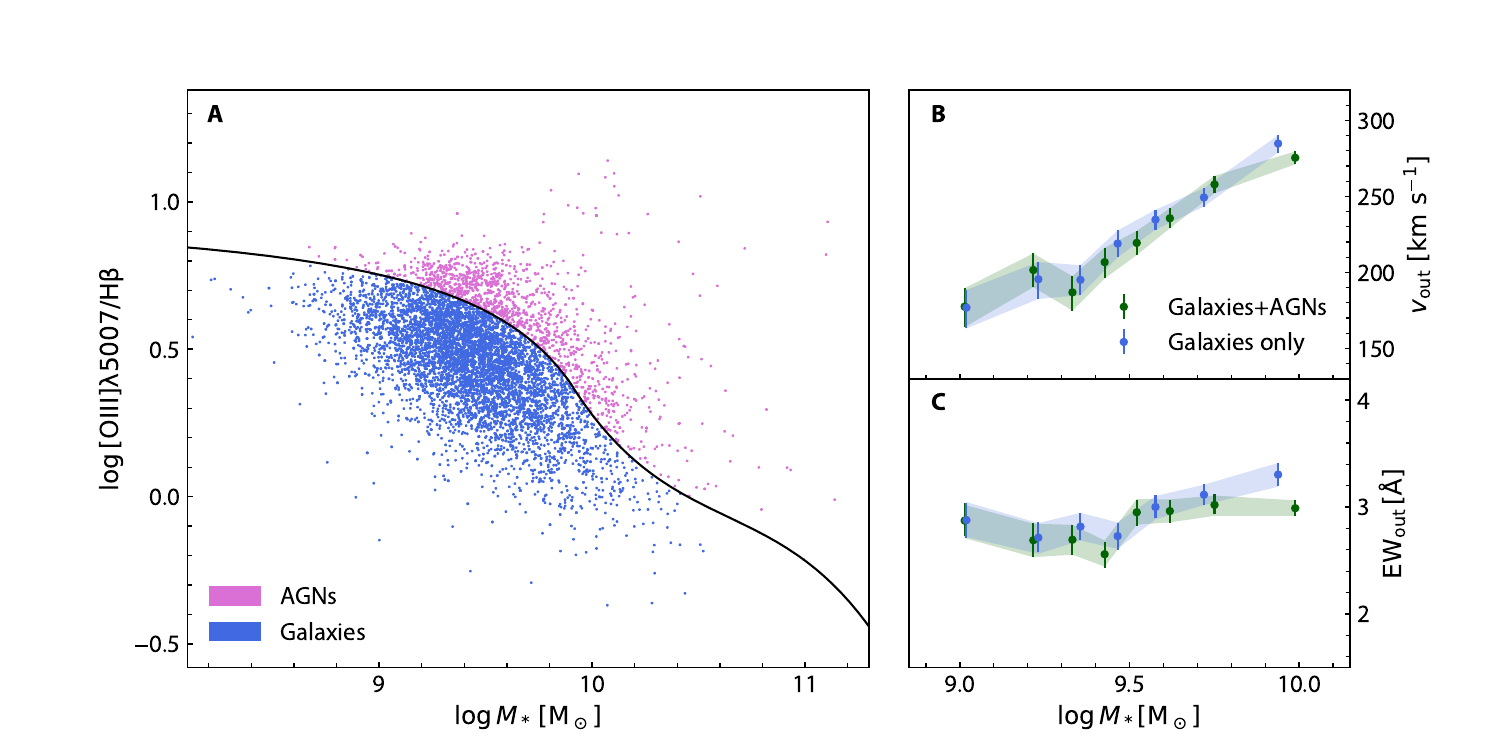}
    \caption{\textbf{Evaluation of AGN contamination.}
    \textbf{(A)}
    the flux ratio [OIII]/H$\beta$ as a function of stellar mass, where the black curve separates AGNs (represented by pink symbols) from normal star forming galaxies (represented by blue symbols).
    For better visualization, we randomly select 20\% of the samples to display in this panel.
    \textbf{(B)}
    $v_{\rm out}$ as a function of stellar mass.
    The green symbols represents the results before removing AGNs while the blue symbols represents the results after discarding AGNs from the sample.
    \textbf{(C)}
    $\rm EW_{out}$ as a function of stellar mass.
    }
    \label{fig:agn}
\end{figure}

Panel A of Figure~\ref{fig:agn} shows the classification of AGNs using the [O\,\textsc{iii}]/H$\beta$ flux ratio versus stellar mass diagram.
In the MEx diagram, the main empirical division is defined as follows:
\begin{equation}
    y=\left\{\begin{aligned}
    0.37/(x-10.5) + 1\quad {\rm if}~x\leq 9.9,\\
    594.753-167.074x+15.6748x^2-0.491215x^3\quad {\rm otherwise}.
    \end{aligned}\right.
\end{equation}
We then examine whether the outflow properties vary significantly with or without AGNs.
The results are shown in the panel B and C of Figure~\ref{fig:agn}. 
Following the spectral stacking methodology applied in Figure~\ref{fig:evolution}, we sort the sample by stellar mass and generate stacked spectra for bins of 5,000 galaxies each. 
We find that including AGNs results in slight variation in $v_{\rm out}$ and $\rm EW_{out}$ of less than 5\%.
However, at $M_*\sim 10^{10}\,\rm M_\odot$, excluding potential AGNs leads to an increase of $\rm EW_{out}$ of $\sim 10\%$, resulting in a weak positive correlation between $\rm EW_{out}$ and $\log M_*$.
This may be due to the contribution of broad Mg\,\textsc{ii} emission lines from AGNs, which can dilute absorption features in the median-stacking analysis.
Consequently, the values of $\rm EW_{out}$, $\dot{M}_{\rm out}$ and $\eta$ at $M_*\gtrsim10^{10}\,\rm M_\odot$ could be underestimated to some extent, which should be treated with caution.

\newpage

\begin{figure}
    \centering
    \includegraphics[width=\textwidth]{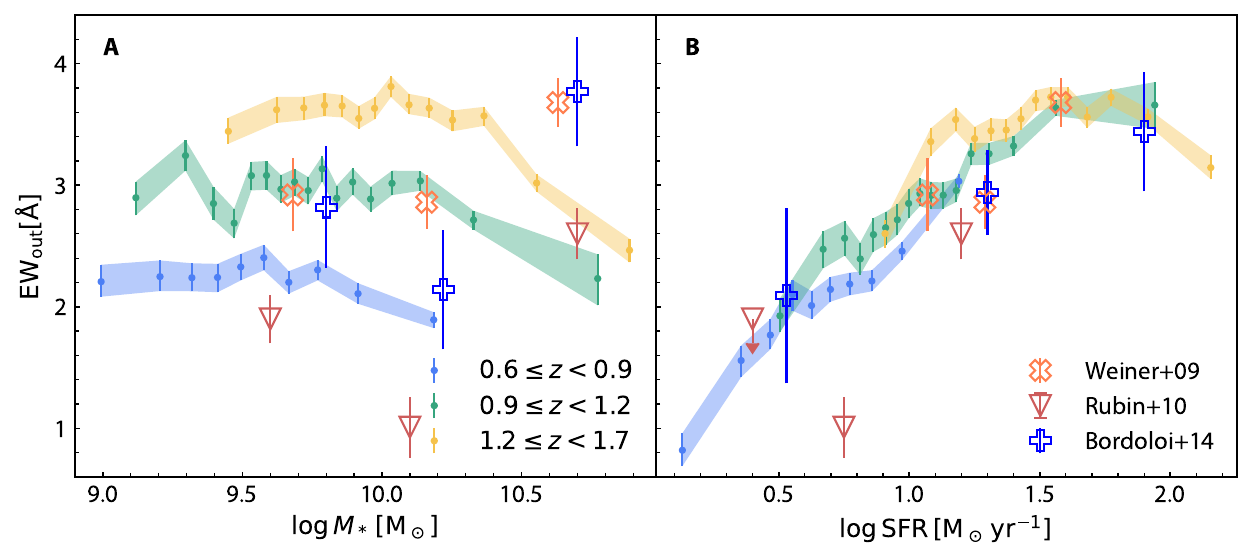}
    \caption{
    \textbf{
    $\rm EW_{out}$ measurements in comparison with previous studies.
    }
    \textbf{(A)}
    The measured outflow equivalent width ($\rm EW_{out}$) in $\log M_*$ bins.
    The data points colored by redshift represent the same bins as those shown in Figure~\ref{fig:evolution} in the main part.
    The outflow EW is estimated using Equation~\ref{eq:out}.
    \textbf{(B)}
    The measured outflow equivalent width ($\rm EW_{out}$) in $\log \rm SFR$ bins.
    The open symbols represent outflow EW measurements obtained from the literature, as indicated in the legend.
    }
    \label{fig:ew}
\end{figure}

\end{document}